\documentclass[5p]{elsarticle}
\usepackage{amsmath,amsfonts}

\usepackage{booktabs} 
\usepackage{algorithmicx}
\usepackage{algpseudocode}
\usepackage{algorithm}
\usepackage{listings}
\usepackage{enumitem}
\usepackage{hyphenat}
\usepackage{comment}
\usepackage{hyperref}

\newcommand{\algoref}[1]{algorithm~\ref{#1}}
\newcommand{\figref}[1]{figure~\ref{#1}}
\newcommand{\Figref}[1]{Figure~\ref{#1}}
\newcommand{\tabref}[1]{table~\ref{#1}}
\newcommand{\eqnref}[1]{equation~\ref{#1}}
\newcommand{\Eqnref}[1]{Equation~\ref{#1}}

\begin{document}

\title{Pushing Memory Bandwidth Limitations Through Efficient Implementations of Block-Krylov Space Solvers on GPUs}

 \author{M. A. Clark}
 \address{NVIDIA Corporation, 2701 San Tomas Expressway, Santa Clara, CA 91214,USA}
\ead{mclark@nvidia.com}

\author{Alexei Strelchenko}
 \address{Fermi National Accelerator Laboratory, Batavia, IL 60510-5011, USA}
 \ead{astrel@fnal.gov}

 \author{Alejandro Vaquero}
 \address{Department of Physics and Astronomy, University of Utah, Salt Lake City, UT 84112, USA}

 \ead{alexvaq@physics.utah.edu}

\cortext[cor1]{Corresponding author}
 \author{Mathias Wagner\corref{cor1}}
\address{NVIDIA GmbH, Adenauer Str. 20A4, 52146 W\"urselen, Germany}
 \ead{mathiasw@nvidia.com}

 \author{Evan Weinberg}
\address{Department of Physics, Boston University, Boston, MA 02215, USA}

 \ead{weinbe2@bu.edu}

\begin{abstract}
The cost of the iterative solution of a sparse matrix-vector system against multiple vectors is a common challenge within scientific computing. A tremendous number of algorithmic advances, such as eigenvector deflation and domain-specific multi-grid algorithms, have been ubiquitously beneficial in reducing this cost. However, they do not address the intrinsic memory-bandwidth
constraints of the matrix-vector operation dominating iterative
solvers. Batching this operation for multiple vectors and exploiting
cache and register blocking can yield a super-linear speed up.
Block-Krylov solvers can naturally take advantage of such batched
matrix-vector operations, further reducing the iterations to solution
by sharing the Krylov space between solves. Practical
implementations typically suffer from the quadratic scaling in the
number of vector-vector operations. We present
an implementation of the block Conjugate Gradient algorithm on NVIDIA GPUs which reduces
the memory-bandwidth complexity of vector-vector operations from
quadratic to linear. As a representative case, we consider the domain of lattice quantum chromodynamics and present results for one of the fermion discretizations. Using the QUDA library as a framework, we demonstrate a 5\(\times\) speedup compared to highly-optimized independent
Krylov solves on NVIDIA's SaturnV cluster.
\end{abstract}


\maketitle


\section{Introduction}
\label{sec:intro}


Two trends in HPC architectures present particular challenges to
scientific software. The wider and wider architectures, in
particular GPUs and Intel Xeon Phi processors, require an increasing
amount of parallelism that cannot always be extracted from algorithms
that had been considered state-of-the-art only a few years ago. Furthermore,
the available memory bandwidth presents itself as the limiting factor 
in most scientific calculations. Hence, in addition to
extracting parallelism exploiting data locality is key to achieve high
performance. This has triggered a lot of algorithmic research, and the
revisiting of older ideas can provide algorithmic speedup and high
efficiency on current architectures even if these ideas had been discarded as
less efficient decades ago.

An important example of an older idea, and the one of focus in this work, is the block-Krylov
solver method. This method reduces the iteration count by processing
blocks of right hand sides (rhs) simultaneously, augmenting the Krylov space
of each rhs by information from the other rhs in the block. On its own, this idea naturally generates more parallelism and exploits data-locality with the simultaneous application of a matrix-vector operator per algorithm iteration.

As will be discussed in more depth later, the original formulation of block-Krylov methods suffered from numerical instabilities~\cite{Nikishin:1995sy} due to overlapping Krylov subspaces and a quadratic scaling in the number of BLAS-like operators with the number of rhs. Later algorithmic developments efficiently overcame the issue of numerical instabilities~\cite{Dubrulle2001}, but the issue of quadratic scaling in operations still rendered the method largely unfeasable. 

In the modern era, this quadratic scaling of operations in a problem featuring a large amount of spatial locality becomes a benefit. Block-Krylov methods naturally provide the arithmetic intensity required to offset memory bandwidth constraints. An efficient implementation offers a significant speedup on top of the multiplicative speedup offered by the features of the algorithm alone. Such algorithms are very suitable for current and future HPC systems, and will continue to contribute
meaningful speedup and develop increased prominence as we trend towards the exascale and beyond.

GPUs provide the perfect setting for demonstrating the block-solver
algorithm, and as such will be the architecture of choice for the study performed here.  Present GPUs typically feature thousands of floating point
units coupled to a very wide and fast memory bus, and, as common to
many-core processors, a high algorithmic arithmetic intensity is
required to reach a reasonable percentage of peak performance. In
this work we utilize the QUDA library~\cite{Clark:2009wm} to implement a high-efficiency
block CG solver that is able to exploit the growing imbalance
between floating-point peak performance and memory bandwidth.

QUDA is a domain-specific library which targets lattice quantum chromodynamics (LQCD), the numerical simulations of the theory of the strong force that
underpins nuclear and particle physics. In the past few years, LQCD
applications have been continually under intensive development to
address computationally demanding scientific problems in nuclear and
high energy physics. These algorithmic improvements and the advances in
computing architectures now allow predictions with errors comparable
or even below experimental results obtained at large particle
colliders like the LHC at CERN.  To control the statistical error, a
typical computational scenario requires the solution of many systems
of linear equations with the same matrix but different rhs. The development of efficient implementations of block-Krylov methods serves as an immediate win in the LQCD community. 

We emphasize that our target problems within LQCD are a representative case of an algorithmic problem present across the discipline of scientific computing. We will not attempt an exhaustive list of fields where the iterative solution of a sparse matrix-vector system against multiple rhs is a predominant use of computer cycles. As a subset of examples, though, we point the interested reader to literature concerning the simulation
of quasistatic electromagnetic fields \cite{2008JCoAM.215..328C}, electromagnetic 3D
modeling \cite{doi:10.1093/gji/ggv216}, scattering and radiation \cite{7857736}, and
fluid simulations \cite{2017arXiv171110622K,Wang:2014:SIN:2663510.2663522}, including aircraft modeling
and weather forecasts. Again, each of these fields can benefit from the efficient implementation of a block-Krylov method described here.

As a last note of importance, the developments in this work are in large part orthogonal and complimentary to existing approaches to reduce the time-to-solution of the problem of solving multi-rhs systems. Block-Krylov methods are compatible with deflation methods, where low eigenvectors of
the linear system are computed and projected out which provides an acceleration in its own right (see, for example,~\cite{eigcg, Wilcox:2007ei}). Block methods are also relevant for components within adaptive multigrid~\cite{Brannick:2007ue, Babich:2010qb} and the
related inexact deflation~\cite{Luscher:2007se}, which are
mathematically optimal (no condition number dependence and constant
iteration count with volume). That being said, we will not discuss these complimentary approaches further.

Our implementation of a block CG solver is, to the best of our knowledge, the first efficient implementation\footnote{Efficient in the sense that we obtain a non-negligible speedup compared to a highly optimized baseline on our target architecture.} of a reliable block solver at scale on GPUs. Previous implementations either ran on CPU~\cite{Calendra2012,Jolivet2016} or didn't provide scaling results~\cite{Bavier:2012:ABD:2590240.2590243,Ji:2014:IBC:2689665.2689675,Ji2017}. We take advantage of mixed precision to mitigate memory-bandwidth limitations and -- as a novel development --  implement a first generalization of the idea of reliable updates~\cite{vanderVorst} to block solvers, a necessary step to correct for round-off errors due to the use of mixed precision. We employ block BLAS operations to saturate memory bandwidth and exploit data locality for both streaming and reduction BLAS operations, allowing us to take advantage of the extra compute power offered by modern nodes. Last, we achieve good strong scaling both by an improved implementation of a multi-right-hand-side-stencil operation and by a new interface which allows us to overlap communications inherent to stencil operations with independent computation.

This paper is organized as follows: in
\(\S\)\ref{sec:previouswork} we highlight previous work in this area,
in \(\S\)\ref{sec:lqcd} we give an overview of the LQCD
computations, and we give an overview of the block CG solver in
\(\S\)\ref{sec:blockcg}.  We introduce the QUDA library in
\(\S\)\ref{sec:quda} and describe in the detail the optimization
techniques employed in our block CG implementation in
\(\S\)\ref{sec:blockcg_impl}.  We show weak scaling performance curves from one and two NVIDIA Quadro GP100 GPUs, and more significantly, strong-scaling performance
curves from the NVIDIA SaturnV DGX-1 supercomputer in
\(\S\)\ref{sec:results}; we discuss future implications in
\(\S\)\ref{sec:future} before finally concluding with
\(\S\)\ref{sec:conclusion}.


\section{Previous Work}
\label{sec:previouswork}

The first use of GPUs for LQCD was reported more than a decade
ago~\cite{Egri2007631}, before the advent of compute APIs,
necessitating the use of graphics APIs.  Since then LQCD has been at
the forefront of the adoption of GPUs in HPC. Notable publications
include multi-GPU parallelization
\cite{Babich:2010:PQL:1884643.1884695, Shi:2011ipdps,
  Alexandru:2011sc}, the use of additive-Schwarz preconditioning to
improve strong scaling \cite{Babich:2011np}, implementation of
multi-grid solvers~\cite{Clark:2016:ALQ:3014904.3014995},
software-managed cache-blocking strategies \cite{inpar2012}, and
JIT-compilation to enable the offload of the entire underlying
data-parallel framework of the Chroma \cite{Edwards:2004sx}
application without any high-level source changes
\cite{Winter:2014dka}.

Block-Krylov-space methods go back to the
1980s~\cite{oLeary1980block} for CG and Bi-CG solvers. By solving
multiple rhs in parallel they combine the benefit of increased
parallelism and data locality with an reduced iteration count compared
to a non-block method. The na\"{i}ve implementation was found to
suffer from stability issues and the subsequent revisions which
introduce explicit re-orthogonalization~\cite{Dubrulle2001} have
significantly improved their general usability.  However, the
stability improvements come at the cost of implementation efficiency
and can become prohibitively expensive outweighing any benefit over
non-block solvers.  In LQCD, block methods have not seen wide
adoption. The use of block BiCGSTAB has been reported
in~\cite{Sakurai2010113,Tadano:2009gg,Nakamura:2011my}. Birk and Frommer generalized
block methods also for the use in shifted systems
\cite{Birk:2012tn,Birk:2014:DCG:2684751.2684768}, which dominate the
runtime in Monte Carlo ensemble generation in LQCD.

To our best knowledge block-Krylov methods have not been reported to
be efficiently implemented for LQCD on GPUs. The benefit of using the
block idea for the matrix-vector operation to multiple rhs has been
reported on GPU and Intel Xeon
Phi~\cite{Kaczmarek:2014mga,Kaczmarek:2014mla}, however there it was
combined with a regular CG solver resulting in only a pseudo-block algorithm.


\section{Lattice Quantum Chromodynamics}
\label{sec:lqcd}

\subsection{Computational Method and Theoretical Relevance}

LQCD calculations are typically Monte\hyp{}Carlo evaluations of
Euclidean\hyp{}time path integrals.  A sequence of configurations of the
{\it gauge fields} \(U\) is generated in a process known as {\em configuration
  generation}. The gauge configurations are importance-sampled with
respect to the lattice action and represent a snapshot of the QCD
vacuum.  Once the field configurations have been generated, one moves
on to the second stage of the calculation, known as {\em analysis}. In
this phase, observables of interest are evaluated on the gauge configurations in the ensemble, and the
results are then averaged appropriately, to form {\em ensemble
  averaged} quantities.  The analysis phase can be task parallelized
over the available configurations in an ensemble and is thus extremely
suitable for capacity-level work on clusters, though calculations on the largest ensembles
can also make highly effective use of capability-sized
partitions of leadership supercomputers.  These calculations rely
heavily on computational resources, with LQCD workloads accounting for
a significant fraction of supercomputing cycles consumed worldwide.

The numerical calculation we will focus on in this study is that of the fermion magnetic
moment. Abscent the contributions from QCD, this quantity can be computed within the standard model
to high accuracy.  The Dirac equation predicts ${\bf \mu} = g e / (2m) {\bf S}$,
where ${\bf \mu}$ is the magnetic moment, $e$ is the fermion electric
charge, $m$ is its mass, ${\bf S}$ is the spin and
$g$ is the Land\'{e} $g$-factor. For free fermions $g=2$, but when we
introduce electro-weak and strong force (QCD) interactions, $g$ receives
perturbative and non-perturbative corrections, which are gathered in
the \emph{anomalous magnetic moment}, defined as \(a_\mu =
\frac{g-2}{2}\), which measures how far we are from the free fermion
case. The particular case of the muon is extremely interesting,
because there is a $3\sigma$ difference between the theoretical and
experimental determinations of $a_\mu$. It is of the utmost importance
to improve the precision of both determinations, for this difference
can be an indicative of new {\it fundamental} physics.  The largest
part of the error in the theoretical determination of $a_\mu$ comes
from non-perturbative QCD contributions, thus the relevance of our target numerical calculation.  The LQCD calculation of
this contribution is essentially computing the trace of the inverse of a large
sparse matrix, the exact calculation of which is unfeasible.
Therefore the calculation relies on stochastic estimators
\cite{Bitar:1988bb}, involving solving thousands of rhs per linear system for many different linear systems. At present, the cost of solving a requisite number of linear systems accounts for~\(\sim90\%\) of the total
computational time.



\subsection{Dirac PDE Discretization}

The fundamental interactions of QCD are encoded in the quark-gluon
interaction differential operator known as the Dirac operator.  As is
common in PDE solvers, the derivatives are replaced by finite
differences.  Thus on the lattice, the Dirac operator becomes a large
sparse matrix, \(M\), and the calculation of quark physics is
essentially reduced to many solutions to systems of linear equations
given by
\begin{equation}
Mx = b.
\label{eq:linear}
\end{equation}
A small handful of discretizations are in common use, differing in
their theoretical properties.  Here we focus on the so-called
staggered form~\cite{Kogut1975}, which is a central-difference discretization; where
the infamous fermion doublers (which arise due to the instability in
the central difference approximation) are removed through
``staggering'' the underlying fermionic (spin) degrees of freedom onto
neighboring lattice sites.  This essentially reduces the number of
spin degrees of freedom per site from four to one, which reduces the
computational burden significantly.  This transformation, however,
comes at the expense of increased discretization errors, and breaks
the so-called quark-flavor symmetry.  To reduce these discretization
errors, the gauge field that connects nearest-neighboring sites on the
lattice is smeared, which essentially is a local averaging of the
field.  There are many prescriptions for this averaging; and here we
employ the popular HISQ procedure \cite{Follana:2003fe}.  When acting
in a vector space that is the tensor product of a four-dimensional
discretized Euclidean spacetime (the lattice) and {\it color} space, the HISQ stencil
operator is given by
\begin{align}
  M_{x,x'} &= - \frac{1}{2} \displaystyle \sum_{\mu=1}^{4} \bigl(
  \hat{U}_x^\mu\, \delta_{x+\hat\mu,x'}\, + \hat{U}_{x-\hat\mu}^{\mu \dagger}\, \delta_{x-\hat\mu,x'} +\nonumber\\
&  \quad\quad\quad\quad \check{U}_x^\mu\, \delta_{x+3\hat\mu,x'}\, + \check{U}_{x-3\hat\mu}^{\mu \dagger}\, \delta_{x-3\hat\mu,x'}\bigr) + m\delta_{x,x'} 
\label{eq:Mstag}
\end{align}
Here \(\delta_{x,y}\) is the Kronecker delta; \(\hat{U}\)
(\(\check{U}\)) is the ``fat'' (``long'') matrix field that connects
lattice sites that are one hop (three hops) apart.\footnote{We adopt the physics convention where \(\dagger\)
  signifies Hermitian conjugation.} These are fields
of $3\times 3$ matrices acting in {\it color} space that live between
the spacetime sites (and hence are referred to as link matrices); and
\(m\) is the quark mass parameter.  The indices \(x\) and \(x'\) are
spacetime indices (the color indices have been suppressed for
brevity).  This matrix acts on a vector consisting of a complex-valued
three-component \emph{color-vector} for each point in spacetime.  When
represented as a sparse matrix, there are 16 off-diagonal matrices,
representing the 1-hop and 3-hop link matrices.  Given the stencil coefficients are
dependent on the underlying gauge field, the coefficients vary for
each configuration in the Monte Carlo ensemble.

For the current problem sizes of interest, the linear system has a
rank that ranges from $V\approx 10^6$ for a small lattice volume
(\(32^4\)) up to $V\approx 10^{10}$ for the largest (\(144^3 \times 288\)).
Furthermore, $M$ is typically very ill conditioned: the
highest eigenvalue is of order $O(1)$, but the lowest one is determined by
the quark mass, which introduces a lower bound in the spectrum. The most
physically interesting cases involve very light or massless quarks, and in
that case the lower bound can become as low as $O(1/V)$.
The standard workhorse algorithm employed is the CG
algorithm, where the normal operator \(A = M^\dagger M\) is used since
\(M\) itself is not Hermitian. In our case we work with the Schur decomposed
version of $M$, which is smaller in size and better conditioned than $M$ or $A$,
although it still shares the problems of the original linear system.

\section{Block CG Solver}

\label{sec:blockcg}
The block-CG algorithm was first published
in~\cite{oLeary1980block}. The idea is to apply the CG
solver~\cite{Hestenes:1952} to a block of $N$ rhs. In a broad sense, scalars in the CG
algorithm  become $N \times N$ matrices and for $N=1$ the
algorithm reduces to the regular CG. 



In infinite precision the block-CG algorithm will converge in at most
$L/N$ iterations where $L$ denotes the length of the vectors, or the rank of the matrix $M$. While
the algorithm maintains a simple link to regular CG, it has issues with
numerical stability and will break down if the residual matrix $\left(R^\dagger
R\right)$ becomes singular, i.e. if the Krylov subspaces of the different vectors in
the block overlap.  There have been various attempts to
remedy this, including dynamically monitoring the rank and removing
any linearly dependent columns~\cite{Nikishin:1995sy}.  A more suitable
approach was introduced in~\cite{Dubrulle2001}, where a set of
retooled block-CG methods are discussed. The common idea is to use a
QR decomposition to enforce the full rank of the $P$ and $R$ matrices
and thus avoid any rank deficits and the need to dynamically remove
columns from the algorithm.  We will only discuss the blockCGrQ
variant, which applies the factorization to $R$, as shown in
\algoref{alg:blockcgrq}. This variant has been shown to generally behave the
best in terms of stability and performance \cite{Dubrulle2001}.

\begin{algorithm}
\caption{blockCGrQ: Solve $A^{L\times L} X^{L\times N} = B^{L \times N}$}
\label{alg:blockcgrq}
\begin{algorithmic}[1]
      \State $A \in \mathcal{C}^{L \times L}$; $R, B, X, Q \in \mathcal{C}^{L \times N}$; $C, S, \beta \in \mathcal{C}^{N \times N}$
  
  \Procedure{blockCGrQ}{$X^{L \times N}, B^{L \times N}$}
    \State $R = B - A X$ \label{blockcgrg:Dslash}
    \State $Q C = R$ \label{blockcgrg:qr1} \Comment{QR decomp.}
        \State $S = \mathbb{I}^{N \times N}$
    \State $P = 0$
    \While{not converged} 
      \State $P = Q + P S^\dagger$ \label{blockcgrg:blas1}
      \State $\beta = \left(P^\dagger A P\right)^{-1}$ \label{blockcgrg:Dslash2}
      \State $X = X + P \beta C$ \label{blockcgrg:blas2}
      \State $Q S = Q - A P \beta$ \label{blockcgrg:qr2}\Comment{QR decomp.}
      \State $C = S C$ \label{blockcgrq:cupdate}
    \EndWhile
  \EndProcedure
\end{algorithmic}
\end{algorithm}

An important observation is that any block algorithm will naively show
quadratic scaling in the number of BLAS vector-vector operations and
reductions.  E.g., the BLAS-1 operations from CG become BLAS-3
operations.  Moreover, the factorization required for stability
increases the cost per iteration. We will address these concerns when we
discuss our implementation in \S\ref{sec:blockcg_impl}.

\section{The QUDA Library}
\label{sec:quda}

QUDA (QCD on CUDA) is a library that aims to accelerate LQCD
computations through offload of the most time-consuming components of
an LQCD application to NVIDIA GPUs.  It is a package of optimized CUDA C++
kernels and wrapper code, providing a variety of optimized linear
solvers, as well as other performance critical routines required for
LQCD calculations.  All algorithms can be run distributed on a cluster
of GPUs, using MPI to facilitate inter-GPU communication. On systems
with multiple GPUs that are directly connected with either NVLink or
PCIe, QUDA will take full advantage utilizing direct DMA copies
between GPUs to minimize the communication overhead.  Similarly, on
systems that support GPU Direct RDMA, where the NIC can read/write
directly to the GPU's memory space, QUDA will utilize this to improve
multi-node scaling.  It has been
designed to be easy to interface to existing code bases, and in this
work we exploit this interface to use the popular LQCD application
MILC~\cite{MILC} as a driver.  The QUDA library has attracted a
diverse developer community and is being used in production at U.S.\
national laboratories, as well as in locations in Europe and India.
The latest development version is always available in a
publicly-accessible source code repository~\cite{githubQUDA}.

The general strategy is to assign a single GPU thread to each lattice
site. Each thread is then responsible for all memory traffic and
operations required to update that site on the lattice given the
stencil operator.  Since the computation always takes place on a
regular grid, a {\it matrix-free} approach is used. Maximum
memory bandwidth is obtained by reordering the vector and link fields
to achieve memory coalescing, e.g., using structures of float2 or
float4 arrays, and using the texture cache where appropriate.
Memory-traffic reduction is employed where possible to overcome the
relatively low arithmetic intensity of the Dirac matrix-vector
operations, which would otherwise limit performance.  The primary
strategy employed here is the use of low-precision data storage,
utilizing single precision or a custom 16-bit fixed-point storage
format (hereon referred to as ``half precision'') together with
mixed-precision linear solvers to achieve high speed with no loss in
accuracy~\cite{Clark:2009wm}.

The library has been designed to allow for maximum flexibility with
respect to algorithm parameter space and maximum performance.  For
example, all lattice objects (fields) maintain their own precision and
data ordering as a dynamic variable.  This allows for run-time policy
tuning of algorithms; these parameters are then bound at kernel launch
time when the appropriate C++ template is instantiated corresponding
to these parameters. 

In fact the use of this autotuning is key to achieving high performance: all kernel launch parameters (block, grid, and shared memory size) are autotuned upon the first call and cached for subsequent reuse. In the work presented here, we extended the autotuner extensively as an algorithm-policy tuner, discussed in \S\ref{sec:dslash} and \S\ref{sec:multireduce}.

\section{Efficient Implementation of BlockCGrQ}
\label{sec:blockcg_impl}

Our implementation of blockCGrQ integrates many developments to overcome the issues discussed in \S\ref{sec:intro} and \S\ref{sec:blockcg}.  We will address these as we step through  \algoref{alg:blockcgrq}
and discuss the details of our efficient implementation. 
We may expect a
reduced time to solution from blockCGrQ due to the reduced iteration
count. An additional benefit can be achieved by applying the
matrix-vector operation to multiple rhs in parallel, exploiting data
locality in this operation; see \S\ref{sec:dslash}. This is countered,
however, by the quadratic increase in the number of BLAS-1
vector-vector operations, which generalize to BLAS-3 matrix-matrix
operations, as well as an additional cost due to the factorization employed
on lines~\ref{blockcgrg:qr1} and~\ref{blockcgrg:qr2}.

We make use of a {\emph{thin QR factorization}\footnote{What makes this thin QR
algorithm more suitable for a large scale, GPU implementation over other more
standard approaches, like the Householder QR algorithm or a modified Gram-Schmidt
factorization, is the large reduction in communications. See for instance
\cite{Fukaya2014,Yamamoto2015}}}~\cite{Golub:1996:MC}. By
strict mathematical definition, the QR factorization of a rectangular
matrix would be given as $\tilde{Q}^{L \times L} \tilde{R}^{L \times
N} = M^{L \times N}$, where $\tilde{Q}$ is dense with orthonormal
columns and $\tilde{R}$ is upper right triangular. A dense matrix of
dimension $L \times L$ is prohibitive in LQCD as $L$ is typically
of $\mathcal{O}(10^{6-9})$. However, the matrix $\tilde{R}$, being upper
right triangular, has zeros in the bottom $(L - N) \times N$
components. Thus, the right $(L - N) \times N$ columns of $\tilde{Q}$
are irrelevant and can be dropped. This leads to the thin QR
factorization given in algorithm~\ref{thinqr}, which of note contains
structured BLAS operations.

\begin{algorithm}
\caption{thin QR: Decompose $Q^{L\times N} R^{N\times N} = M^{L \times N}$}
\label{thinqr}
\begin{algorithmic}[1]
  \Procedure{thinQR}{$M^{L \times N}$}
    \State $H^{N \times N} = \left(M^{L \times N}\right)^\dagger M^{L \times N}$
    \State $\left(R^{N \times N}\right)^\dagger R^{N \times N} = H^{N \times N}$ \Comment{Cholesky decomp.}
    \State $Q^{L \times N} = M^{L \times N} \left(R^{N \times N}\right)^{-1}$ 
  \EndProcedure
\end{algorithmic}
\end{algorithm}

With this exposition, we can now decompose the operations in our implementation of \algoref{alg:blockcgrq} into four classes:
matrix-block-vector operations in lines \ref{blockcgrg:Dslash} and~\ref{blockcgrg:Dslash2};
block-vector-block-vector streaming operations in lines \ref{blockcgrg:blas1} and~\ref{blockcgrg:blas2};
block-vector-block-vector reduction operations in lines \ref{blockcgrg:Dslash} and~\ref{blockcgrg:Dslash2}; and
small dense matrix operations.  The latter three kinds of operations are also embedded in the thin QR decomposition. 

A key ingredient for the implementation is generalizing our lattice fields from a single vector of length $L$ to an efficient data structure of size $L \times N$. We refer to these $L \times N$ matrices as {\emph{composite fields}} with details discussed in \S\ref{sec:compositefields}.  We will discuss how to exploit
data locality to reuse $A$ in the matrix-vector operation to obtain a
speedup per rhs in \S \ref{sec:dslash} and mitigate the na\"{i}ve
quadratic scaling of the block-vector-block-vector operations to
achieve an overall multiplicative speedup
in \S \ref{sec:multiblas}, \ref{sec:multireduce}.

For all small $N \times N$ matrices we exploit the Eigen
library~\cite{eigen} for the Cholesky decomposition of $R$. Given that
$N\leq64$, these matrices are comparably small and all operations are
negligible in the overall runtime.

\subsubsection*{Mixed-precision implementation}
For our mixed-precision implementation we rely on ``reliable
updates'', referring to a class of residual correction methods
described, e.g., in~\cite{vanderVorst,Sleijpen1996}. A general
residual correction method periodically updates the iterated residual
with the true residual from the iterated solution. This corrects for
numerical round-off effects and, by extension, possible instability in
Krylov methods. In the case of mixed-precision solves, this correction
is often essential to monitor convergence in a meaningful fashion. We
based our implementation of reliable updates on the implementation
given in~\cite{Clark:2009wm}, with important deviations presented in
\algoref{alg:reliableupdates}. We believe that this is the first
implementation of a residual correction method to improve the
convergence of a block-Krylov method.

\begin{algorithm}
\caption{Reliable updates}
\label{alg:reliableupdates}
\begin{algorithmic}[1]
  \Procedure{ReliableUpdate}{$X^{L \times N}, \widehat{Q}^{L \times N}, S^{N \times N}$}
    \State $Y^{L \times N} = Y^{L \times N} + X^{L \times N}$ \Comment{\texttt{caxpy\_5d}}
    \State $X^{L \times N} = 0$
    \State $R^{L \times N} = B^{L \times N} - A^{L \times L} Y^{L \times N}$ \Comment{Block Mat-Vec, \texttt{caxpy\_5d}}
    \State $\widehat{Q}^{L \times N} C^{L \times N} = R^{L \times N}$ \Comment{thin QR}
    \State $S^{N \times N} = C^{N \times N} \left(C_{old}^{N \times N}\right)^{-1}$ \label{reliableupdates:supdate}
  \EndProcedure
\end{algorithmic}
\end{algorithm}

Like in~\cite{Clark:2009wm}, we employ a persistent fine precision accumulator for
the iterated solution, denoted by the block vector $Y^{L \times
N}$. The block vector $X^{L \times N}$ only contains the iterated
update to the solution between reliable updates, which, in contrast
to~\cite{Clark:2009wm}, is maintained at full precision. The block vectors $P^{L\times N}$ and $Q^{L \times N}$ are maintained in lower precision. We perform a
reliable update when the {\emph{maximum relative}} iterated residual decreases
by a factor of $\delta$. We use the relative residual as opposed to
the absolute residual to perform a meaningful comparison between the
different rhs vectors. The iterated residual can be computed after $C^{N \times N}$ is updated on line~\ref{blockcgrq:cupdate} of \algoref{alg:blockcgrq} by the identity $\left|R^{L \times N}_i\right|^2 = \sum_{j} \left|C_{ij}^{N \times N}\right|^2$. 

When a reliable update is triggered, the current iterated solution update 
$X^{L \times N}$ is accumulated into the full solution $Y^{L \times N}$ and the true
residual is computed in full precision, as implemented
in~\cite{Clark:2009wm}. As required by blockCGrQ, we perform a thin QR
factorization of the true residual. The update of $S^{N \times N}$ on
line~\ref{reliableupdates:supdate} of algorithm~\ref{alg:reliableupdates} forces the preservation of
the identity given in equation (6.1) of~\cite{Dubrulle2001}. The new
$S^{N \times N}$ enters the update of $P^{N \times
N}$ on line~\ref{blockcgrg:blas1} of algorithm~\ref{alg:blockcgrq}. This is
essential to maintain one of the defining properties of blockCGrQ,
$\left(Q^{L \times N}_{old}\right)^\dagger P^{L \times N}
= \mathbb{I}^{N \times N}$, noted below figure 6.1
of~\cite{Dubrulle2001}, which is intimately connected to preserving the Krylov search space through the reliable update.

\subsection{Composite Fields}
\label{sec:compositefields}

In order to provide both ease of expressibility and high efficiency
for the block-solver implementation, we extended QUDA to facilitate
the application of the required linear algebra components on both sets
of vector fields and individual vector fields.  To this end we
enhanced QUDA's vector-field container {\it ColorSpinorField}, with a
new attribute: whether it is a {\it composite} field or not.
When it is a composite field, the container contains an STL vector of
{\it component} ColorSpinorFields, whose memory refers to contiguous
subsets of the parent composite.  The composite field can thus be
viewed as a single higher-dimensional object while providing accessor methods
to expose individual components or the entire STL vector.
This functionality of
being able to switch between block and scalar methods while using the same data structure
was critical in algorithm prototyping, debugging, and performance analysis.

%

\subsection{Block-optimized Matrix-vector Product}

\label{sec:dslash}

The efficient application of the stencil operator, \eqnref{eq:Mstag}, is critical in a performance-optimized iterative
linear solver.  The stencil connects lattice sites that are one and
three hops away in all dimensions in both positive and negative
directions relative to the center site.  To apply the stencil, we load the
neighboring sites (6 numbers per site), the link matrix between the
neighboring sites (18 numbers per matrix), perform the link-matrix
times vector multiplication and accumulate to the new resulting center
value.  With 16 neighbors, this means each application of the operator
requires 1158 flops and 390 words of memory traffic per lattice
site, for an arithmetic intensity of 0.74 in single precision.  The
Pascal-generation GPUs used in this study can achieve a STREAM
bandwidth of 575 GB/s, giving a na\"{i}ve performance upper bound of
435 GFLOPS in single precision, compared to a peak of 10.1 TFLOPS, so
we can see that any na\"{i}ve implementation of the stencil would be
memory-bandwidth bound.

In this specific algorithm, we apply this stencil to {\it multiple} vector fields
simultaneously, and as such we have
implemented a block-optimized stencil operator which allows us to severely reduce the contribution of link matrices to the overall memory traffic. In the usual single-rhs stencil
operator, a volume of threads are launched, each assigned to a grid
point.  Threads are partitioned into a grid of thread blocks, where
each thread block runs on a given streaming multiprocessor (SM),
roughly equivalent to a CPU core.  This provides enough thread-level
parallelism to saturate the GPU memory bus.

When augmenting the GPU kernel to apply the stencil to multiple sites,
we utilize the Cartesian thread indexing that GPUs expose: we map the
y-thread dimension to the rhs index, keeping the x-dimension thread
mapping to grid points unchanged.  Since all threads in the same thread
block share the same L1 cache, threads with a common grid index in the
same thread block (e.g., different rhs index) are able to exploit
temporal locality to share the link matrix load. The optimal x-y shape
of the thread block is autotuned: this effectively optimizes for the
competing localities (link-matrix element reuse versus vector-field neighbor
reuse) and L1 vs L2 reuse (larger blocks will result in more L1 cache
hits versus L2 cache hits).

\begin{figure}
  \includegraphics[width=.4\textwidth]{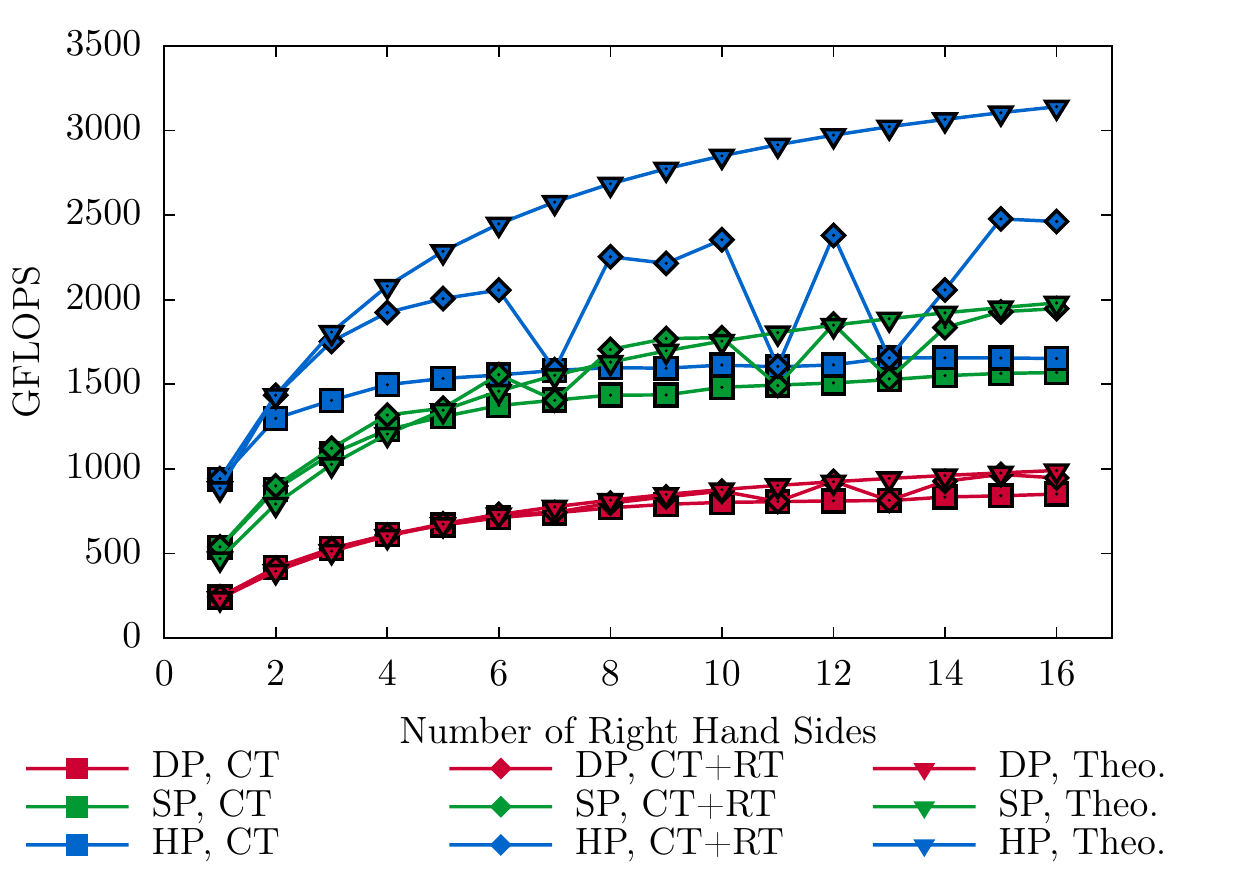}
\caption{\label{fig:dslash} The performance in GFLOPS of the stencil operator as a function of number of rhs vectors. Measurements are taken from a $24^4$ lattice. The colors red, green, and blue correspond to double, single, and half precision, respectively. The solid lines with filled squares represent the cache-tiling implementation, the dashed lines with filled diamonds the cache-and-register-tiling implementation, and the dotted lines with filled triangles is a roofline model of the performance. Tests were performed on a Quadro GP100 GPU.}
\end{figure}

In \figref{fig:dslash} we plot the performance results from the
block-optimized stencil for a $24^4$ lattice. We use the colors red, green, and blue to represent double, single, and half precision, respectively. As an alternative classifier, the bottom, middle, and top batches of lines correspond to double, single, and half precision, prespectively. The solid lines with filled squares represent the performance of
the cache-tiled implementation for double, single and half precision.
While we see an improvement in performance with increasing rhs, the
improvement for single precision, and in particular half precision is
milder.  Further analysis revealed the limiting factor to be
load-instruction rate and cache bandwidth.  To alleviate this we
utilize a {\it register-tiling} approach where each thread is
responsible for updating multiple rhs at constant grid index.  The
register-tile size is a template parameter in our code, with tile
sizes \(1\ldots6\) instantiated.  The final algorithm is a hybrid of
cache and register tiling, where the optimal balance is found by the
autotuner.  The performance of the hybrid approach is shown in the
dashed lines with filled diamonds, where we see minimal improvement for double performance,
but significant improvement for both single and half precision. The
erratic behavior at larger numbers of rhs is due to numbers of rhs that
are not divisible by a tile size greater than 1 or 2.  We also include
a roofline extrapolation of the expected performance from exploiting
the temporal locality of the link-matrix elements.  Here we are also
assuming perfect reuse of vector-field loads in the x dimension, and
50\% in the y dimension (e.g., for every 16 grid points each thread
loads, 6 of them are cached).  For double and single precision, the
hybrid tiling performance broadly matches the roofline expectation, given as the dotted line with filled triangles,
with the roofline actually underestimating performance at small rhs in
single precision: this is due to the cache-hit rate for the
vector-field loads being greater than in the model.  For half
precision, even with hybrid tiling, the kernel remains
load-instruction-rate bound, as evidenced by it failing to match the
roofline performance model.

The multi-GPU parallelization of stencil operators in QUDA has been
reported
elsewhere~\cite{Babich:2010:PQL:1884643.1884695,Babich:2011np}.  The
essential strategy is to overlap the communication of the halo region
with the computation on the interior, with a final halo update kernel
required to produce the complete answer.  The ability to hide the
communication scales with the surface-to-volume ratio, hence strong
scaling is often communication limited.  To maximize performance we
made use of two key technologies:
\begin{itemize}[leftmargin=0.2in]
\item Peer-to-peer communication: direct communication between GPUs within the same node
\item GPU Direct RDMA: direct communication between the GPUs and the NICs maximizing the inter-node communication rate
\end{itemize}

On a sufficiently distributed system, the communication of the halo
region can take much longer than the computation on the interior. To help mitigate this, we
have implemented an interface to the stencil kernel, such that an algorithm can schedule additional
computation during the halo exchange that is independent of the stencil application. This is an important optimization for a wide class
of algorithms, but in particular for the block-Krylov method discussed
here since the amount of work scales with the number of right hand
sides.

We note that on line~\ref{blockcgrg:blas2} of \algoref{alg:blockcgrq}, the update of the iterated $X$ can be scheduled to occur at any point after the calculation of $\beta$. We thus move the update of $X$ to occur during the stencil application in the next iteration. This depends on maintaining the ``old'' version of $P$ from before the update on line~\ref{blockcgrg:blas1}. Our implementation of line~\ref{blockcgrg:blas1} preserves the state of $P$ from when $X$ would normally be updated at no memory traffic overhead---there is no additional cost from updating $X$ during the stencil application. This allows for a simplified interface where the scheduled compute can still occur during the would-be communication on a single node at no additional cost. One exception to this situation is when a reliable update is triggered. In that case, $X$ must be updated immediately. This is infrequent and thus negligible.


\subsection{Streaming Multi-BLAS}
\label{sec:multiblas}
For the block linear algebra operations we distinguish
between block-AXPY-type operations and block-DOT-type operations.
They differ conceptually, and in implementation, since the latter
require global reductions.  We refer to the former
as {\it Streaming Multi-BLAS}, which we focus on here, and the latter as
{\it Reduction Multi-BLAS}, which we delay until \S\ref{sec:multireduce}.  

For the
discussion here we chose a \texttt{caxpy} operation as the representative
example. In blockCGrQ it is used in the form $Y^{L \times N} = Y^{L
  \times N} + X^{L \times M} a^{M \times N}$, or at the element level,
\begin{equation} \label{eq:blockcaxpy}
y_{k,j} = y_{k,j} + \sum_{i = 0}^{N - 1} x_{k,i} a_{i,j}.
\end{equation}
Note that in the case of blockCGrQ we only need the case $N = M$, and as such $a^{N \times N}$ is a square matrix. We will assume $M = N$, the number of right hand sides, unless noted otherwise.

\Eqnref{eq:blockcaxpy} can be implemented as separate
\texttt{caxpy} applications for each component of $\alpha^{N \times
  N}$. This implementation incurs a quadratic scaling of memory
traffic and floating-point operations with $N$.  Due to the low arithmetic
intensity, the operation is severely limited by memory bandwidth and becomes prohibitively
expensive as the number of right hand sides increases. A block
\texttt{caxpy} takes advantage of the memory reuse inherent in
equation~\ref{eq:blockcaxpy} by re-expressing the set of vector-vector
operations as a single matrix-matrix computation. For a given
component $k$, block \texttt{caxpy} loads all $N$ components of $X^{L
  \times N}_k$ and $Y^{L \times N}_k$, performs the $N^2$
multiplications and additions, then saves all $N$ components of $Y^{L
  \times N}_k$ back to memory. The quadratic scaling of memory traffic
becomes a linear scaling.  Since the required flops remains well below
the theoretical peak of modern GPUs, the quadratic cost of the
algorithm shifts to lower latency cache load instructions which can
take advantage of the spatial locality inherent to cache.

We have developed a general framework to perform such BLAS-3-type
operations. We distribute the update of a single block vector
constituent $k$ over $N$ threads on the GPU by assigning a unique
constituent $y_{k,j}$ to each thread. Each thread is tasked with
updating $y_{k,j}$ with the contributions from each $x_{k,i}$.
Similar to the block-matrix-vector product in \(\S\ref{sec:dslash}\),
we use CUDA thread blocks to ensure that for a given $k$, all $j=0,
\ldots,N$ are scheduled to the same SM. This guarantees that all
$x_{k,i}$ reside in L1 cache and can be shared between all $N$
threads, exploiting spatial locality.

We take extensive advantage of \texttt{C++} templating which allows
the compiler to perform additional optimizations. We template over all
possible combinations of the block sizes of $N$ and $M$ up to a
sufficient maximum, enabling the compiler to perform loop
unrolling. In addition, we template over all possible precisions for
$Y^{L \times M}$ and all equal or lower precisions for $X^{L \times
  N}$. This optimization is essential for efficient mixed-precision
block-Krylov methods.  Furthermore we take advantage of QUDA's
autotuner in the developed BLAS-3 framework.  Finally, we note that while
the block caxpy is essentially a CGEMM operation included with every
standard BLAS library, the disparity between the sizes of \(N\) and
\(L\) ($L\gg N$) are such that no out-of-the-box BLAS library would perform well
for this use case, requiring us to implement our own framework to this end.

\begin{figure}
  \includegraphics[width=.4\textwidth]{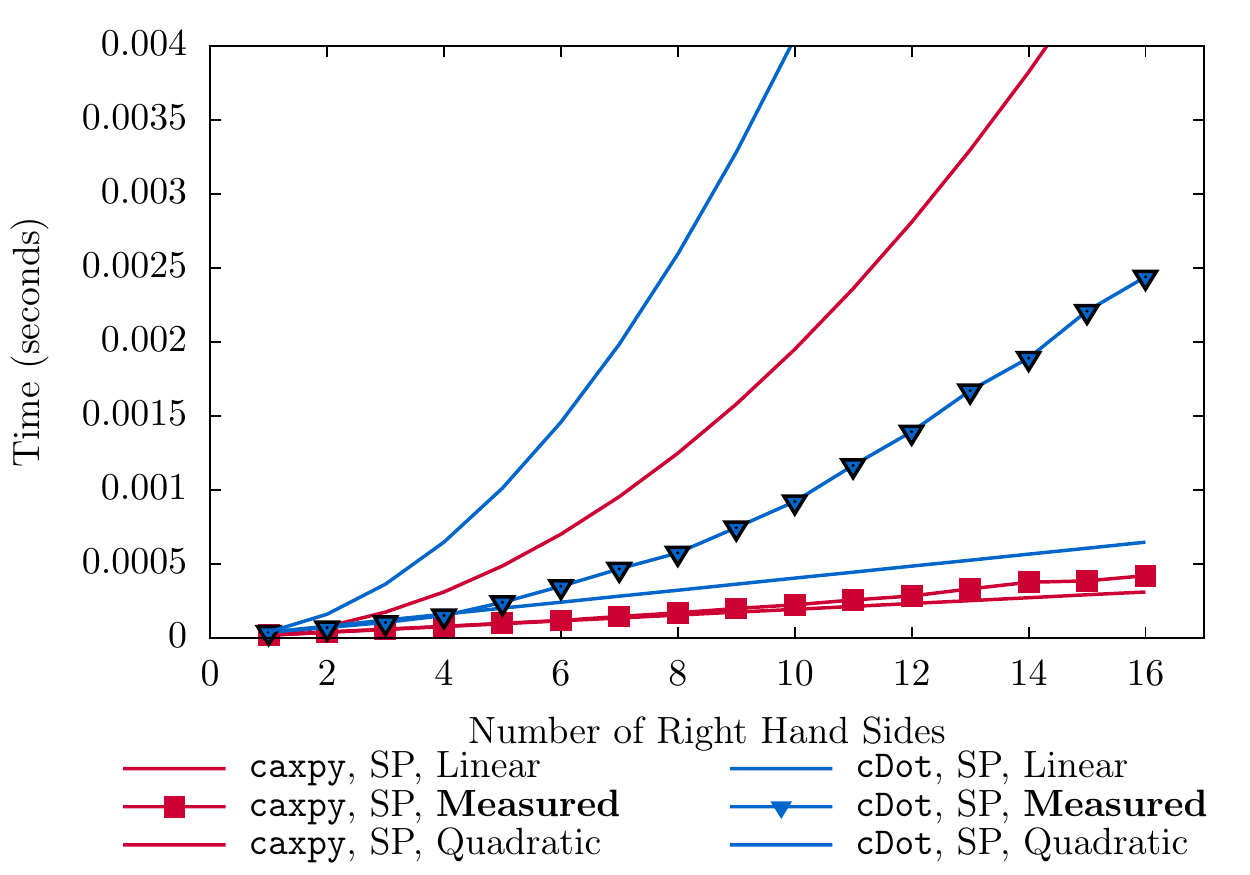}
\caption{\label{fig:blocktime} The average cost, in seconds, of applying a square block \texttt{caxpy} and a square block \texttt{cDotProduct} as a function of the number of right hand sides. Measurements are taken from a $24^4$ lattice, in single precision. Both are contrasted against a best-case linear and a worst-case quadratic extrapolation from a single \texttt{caxpy} and \texttt{cDotProduct}. The block \texttt{caxpy} features linear scaling for a wide range of $N$. In contrast, the block dot product shows quadratic scaling even for small $N$. Tests were performed on a Quadro GP100 GPU.}
\end{figure}

\Figref{fig:blocktime} shows that the cost of the block-caxpy operation scales linearly for a wide range of the number of rhs. The eventual deviation from linear scaling is due to the less-penalizing cache traffic. This is consistent with \figref{fig:blasflops}, where we see that the performance of the block-capxy operations also scales linearly in $N$. Note that due to finite cache sizes there is a turnover point in $N$ where it is advantageous to recursively divide the block BLAS into tiles, e.g., to split a $24 \times 24$ block operation into four $12 \times 12$ block operations. We refer to a recursed segment of the BLAS operation as a ``tile'', and its size as the ``tile size".

There are two places in \algoref{alg:blockcgrq}, lines \ref{blockcgrg:qr1} and \ref{blockcgrg:blas1}, where we can take advantage of recursively divided block-BLAS operations and perform additional optimizations. In each case, we take advantage of the triangular structure of the small, dense matrix $A$ and skip computing empty tiles. The benefit of this optimization scales quadratically in $N$ over tilesize. However, we observe that this optimization is only beneficial simultaneous with when tiling itself becomes beneficial.


\subsection{Reduction Multi-BLAS}
\label{sec:multireduce}
For the discussion of block reductions, we will consider the representative case of a complex dot product, or \texttt{cDotProduct}, $K^{N \times
  M} =\left(X^{L \times N}\right)^\dagger Y^{L \times M}$. At the
component level it is defined as
\begin{equation}
k_{ij} = \sum_{k = 0}^{L-1} x_{k,i}^* y_{k,j}.
\end{equation}
In the following we again limit our discussion to the case $N=M$ and assume that the resulting matrix $K^{N \times N}$ is Hermitian, which is always true in blockCGrQ. 

At a basic level, our software implementation of block reduction
operations is similar to the implementation of Streaming Multi-BLAS
operations. The goal is again to achieve an implementation of Reduction
Multi-BLAS operations that obeys a linear scaling in memory traffic as opposed to a quadratic scaling. We template over
both the left and right block vectors, allowing the compiler to
generate efficient code for every combination of left and right block
sizes and apply the autotuner separately to each combination. Each
{\emph{thread}} is responsible for the multiplication of a single
component $y_{k,j}$ with every component of $x_{k,i}$. We utilize the CUB
library~\cite{cub} for an efficient thread block and inter-block
reduction per site. We still recursively
divide block reduction operations and take advantage of the known
structure of the resulting dense matrices.  Finally, on distributed systems, we only perform the global
MPI reduction once all inner tiles have completed. We emphasize that
an efficient implementation of Reduction Multi-BLAS requires a more
careful recursion strategy when tiling.

\begin{figure}
  \includegraphics[width=.4\textwidth]{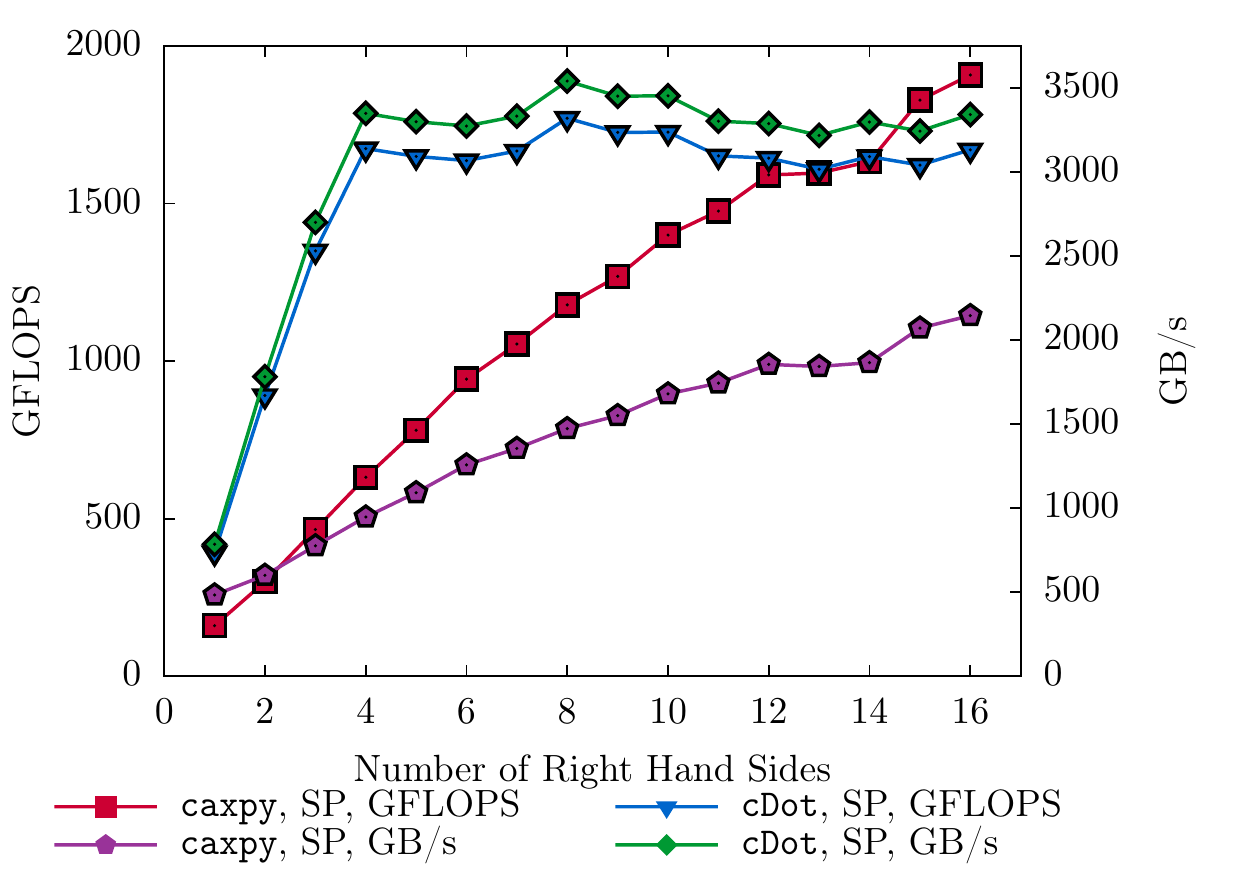}
\caption{\label{fig:blasflops} The GFLOPS (left) and GB/s (right) of block \texttt{caxpy} and \texttt{cDotProduct} as a function of the number of right hand sides. Measurements are taken from a $24^4$ lattice, in single precision. Block \texttt{caxpy} scales nearly linearly in the number of right hand sides, while the block \texttt{cDotProduct} very quickly saturates cache bandwidth. Tests were performed on a Quadro GP100 GPU.}
\end{figure}

We see in \figref{fig:blocktime} that reductions scale linearly in
cost for a {\emph{much}} narrower regime of $N$ than the block
\texttt{caxpy} operation. This is consistent with what is seen in
figure~\ref{fig:blasflops}. Block reductions quickly saturate cache
bandwidth as a function of $N$ and tiling thus becomes essential even
for modest numbers of right hand sides. On top of tuning the internal
block reduction, we tune the tile size for recursive reductions to be
mindful of two different potential optimizations via a {\emph{policy
    class}}. First, we perform distinct tunings for when the left and
right block vectors alias versus when they do not. When the
two block vectors alias, the subset of reductions on the ``block-diagonal'' inner tiles
feature memory-load reuse. Second, we perform distinct tunings for when the
output dense matrix is Hermitian versus when it is not Hermitian.  In
blockCGrQ, the output dense matrices are always Hermitian, thus we
only perform a block reduction on inner tiles (without loss of
generality) above the block diagonal. We then reconstruct the inner
tiles below the block diagonal by taking advantage of the overall
Hermiticity of the matrix.
Smaller tile sizes further reduce the total number of redundant calculations. For this reason, it may be
beneficial to use a smaller tile size despite an overall increase in
memory streaming. This is automatically optimized by policy-class tuning.

We also explored the potential optimization of not computing elements
within the inner-block-diagonal tiles below the inner diagonal. However,
as this only reduces the number of floating point operations in a
tile and does not decrease the overall memory traffic, no benefit was
found.

Tuning the tile size is crucial for fully optimizing block
reductions. The magnitude of the vector size $L$ leads to different
optimal tile sizes. Less threads are used in the case of smaller $L$
and larger tile sizes are needed to saturate memory bandwidth. In
contrast, more threads are used in the case of larger $L$, which can
saturate the memory bandwidth with smaller tile sizes. The optimal
tile size also depends on the precision of the input block
vectors. High-precision input vectors will saturate memory bandwidth
more quickly than lower-precision input vectors. For this reason, high
precision input vectors prefer smaller tile sizes. Low precision input
vectors will instead saturate cache load instructions, which also
leads to a preference towards smaller tile sizes.

\subsection{Benchmarking the implementation}
\label{sec:benchmark}
\begin{figure}
  \includegraphics[width=.4\textwidth]{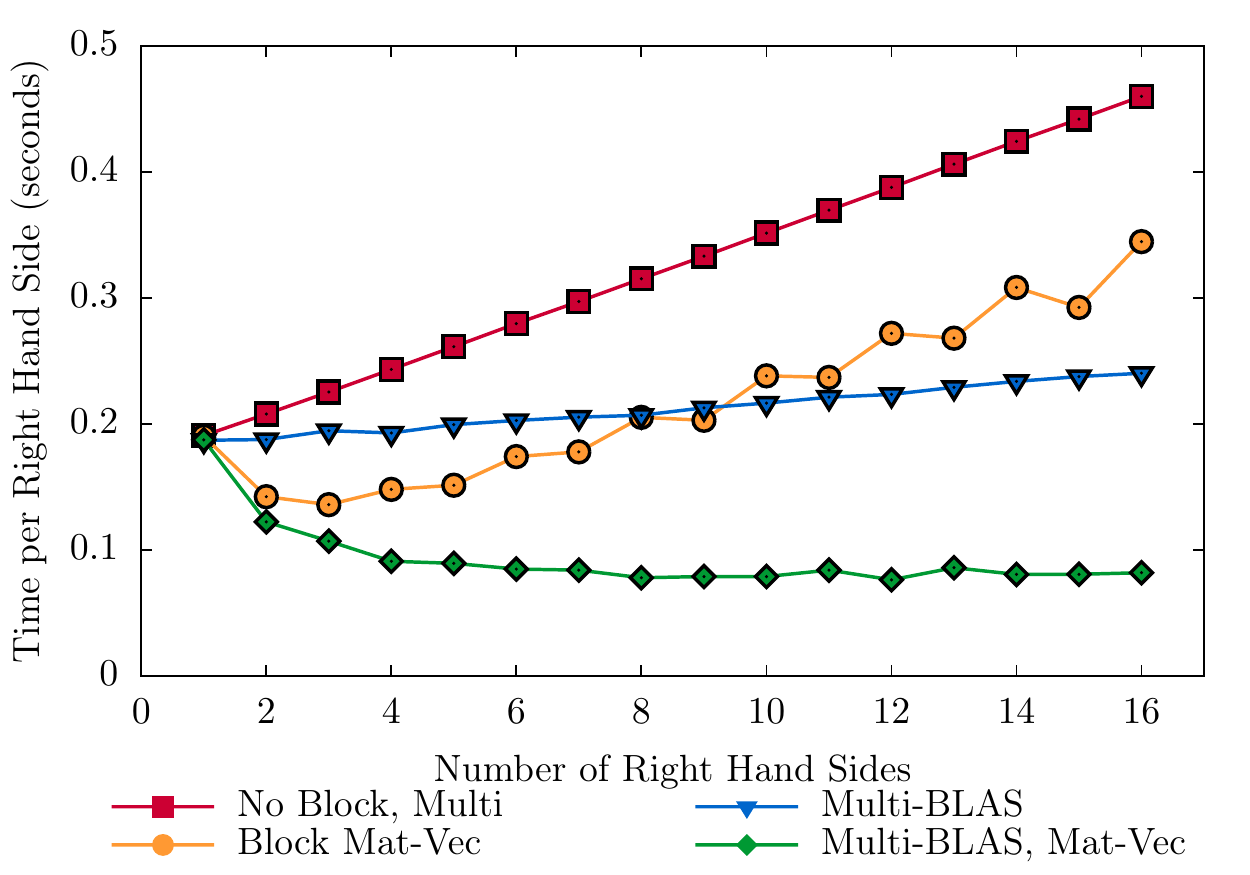}
\caption{\label{fig:timeperrhs}Time for 100 iterations of blockCGrQ. We show the na\"ive implementation and the effect of adding in the optimized block BLAS routines and also the block-optimized matrix-vector product. Tests were performed on a Quadro GP100 GPU.}
\end{figure}

Optimizing the streaming and reduction multi-BLAS kernels to obtain almost linear
scaling is a crucial component in
our implementation. Without this, the lower
iteration count would not translate into a reduced time to
solution. In \figref{fig:timeperrhs} we show the time for running
100 iterations of blockCGrQ in double precision.

For the na\"{i}ve implementation, ``No-Block'', the time per rhs rises
linearly as expected from the quadratic scaling of the sequential BLAS-1
operations. Turning on the corresponding BLAS-3 operations is the most important
optimization. It brings the time per rhs down to almost constant and a
reduced iteration count will almost directly translate to a reduced
time to solution. Switching to the version that also enables the
block-matrix-vector kernel significantly reduces the time per iteration for a
small number of rhs and we see an almost flat behavior between 8 and
16 rhs, at only half of the time of one iteration of a
CG solver with 1 rhs. Hence we can expect to see a multiplicative
speedup from the block-matrix-vector operation and the reduced
iteration count from blockCGrQ.

\section{Results}
\label{sec:results}


\subsection{Methodology}
\label{sec:methodology}

To test the algorithm in the context of the target measurement of the
disconnected part of the contribution to the muon $g-2$, we used
development versions of the MILC code and QUDA.  We compiled using GCC
v4.8.5 and the CUDA-toolkit 8.0. The Small and Medium data sets were
processed on a workstation equipped with two Quadro GP100 GPUs using
MVAPICH2 2.2. The Large and X-Large data sets were run on the NVIDIA
SaturnV cluster: each node consists of a DGX-1 server, with eight
Tesla P100 GPUs with the NVLink intra-node fabric and a quad EDR
Infiniband interconnect between nodes. OpenMPI 1.10.6 was used as the communications
layer. We used gauge configurations from an ongoing $g-2$ project, as
shown in the following table.

\begin{table}[h!]
{\renewcommand{\arraystretch}{0.8}
\begin{tabular}{|l|cccc|}
\hline
Label & $L_s$ & $L_t$ & $a$ (fm) & $m_\pi$ (MeV) \\
\hline
 Small  &   16  &   48  &    0.15  & $\approx 265$   \\
 Medium &   32  &   48  &    0.15  & $\approx 133$   \\
 Large  &   48  &   64  &    0.12  & $\approx 133$   \\
X-Large &   64  &  192  &    0.06  & $\approx 297$   \\
\hline
\end{tabular}
}
\caption{Lattice configurations and their physical parameters;
  the pion mass \(m_\pi\) roughly corresponds to the quark mass parameter that
  appears in \eqnref{eq:Mstag}.}
\label{tab:mpi}
\end{table}

The goal of our tests is to determine if the optimizations explored in \(\S\)\ref{sec:blockcg_impl} bear fruit when combined in blockCGrQ.
As a model test case, we consider stochastic trace estimation for the HISQ
stencil given in \eqnref{eq:Mstag}. This is a relevant measurement for
the $g-2$ project and is also a scenario which necessarily requires a large number
of linear solves. In this case, we have a total of 192 rhs per test: 24 noise sources,
each partitioned into 8 disjoint subsets. For the 1--2 GPU tests on the Small and
Medium datasets, we utilize  the
Truncated Solver Method (TSM) \cite{Bali:2009hu}, which consists of a
predictor-corrector scheme, partitioning the trace of the inverse
matrix into two sweeps: first a cheaper bulk prediction is calculated,
at the expense of introducing a bias that must be corrected in a
subsequent high precision (fine) computation. The sloppy relative residual was set to $5\times 10^{-6}$,
whereas the fine solves used $2\times 10^{-9}$. For simplicity, in our strong scaling tests,
we only consider fine solves to a relative tolerance of $2\times 10^{-12}$. 

The primary quantity of interest is the total time to iteratively solve $M$ against all 192 rhs. Our baseline is the current CG implementation in QUDA where we sequentially solve on all rhs. For comparison, we perform blockCGrQ in blocks of rhs of size 8, 16, 24, 32, 48, and 64, leading to 24, 12, 8, 6, 4, and 3 applications of blockCGrQ, respectively. We skip some cases as applicable due to insufficient memory or unreasonably small local volumes. We consider both the case of full double precision and a mixed double-single solve to explore the viability, stability, and potential benefits of block reliable updates. We choose to trigger a reliable update when the relative residual drops by a factor of 10. Double-half mixed-precision solves will be a point of future work as noted in \(\S\)\ref{sec:future}.

\begin{figure}[t]
\includegraphics[width=0.95\linewidth,angle=0]{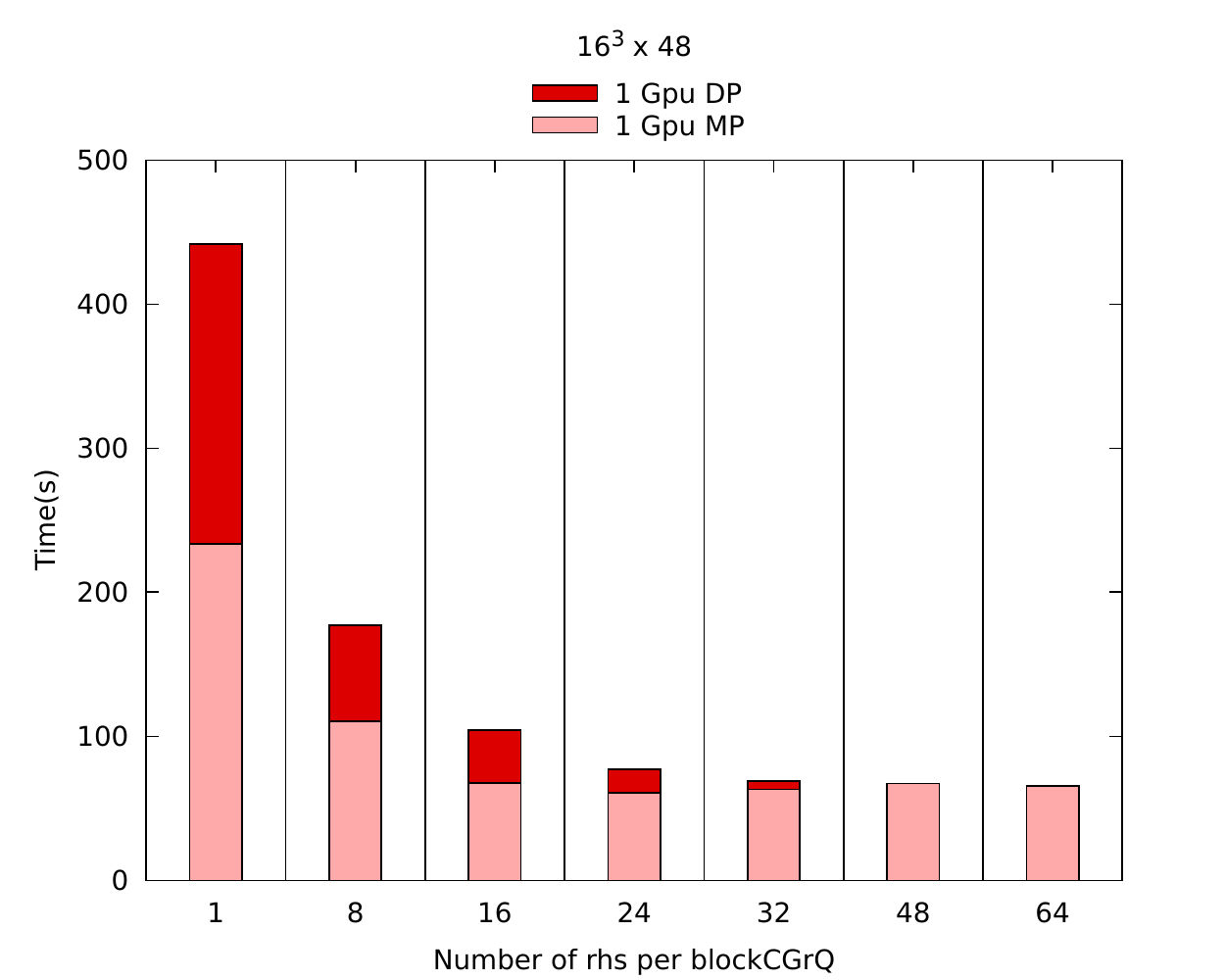}
\\~\\
\includegraphics[width=0.95\linewidth,angle=0]{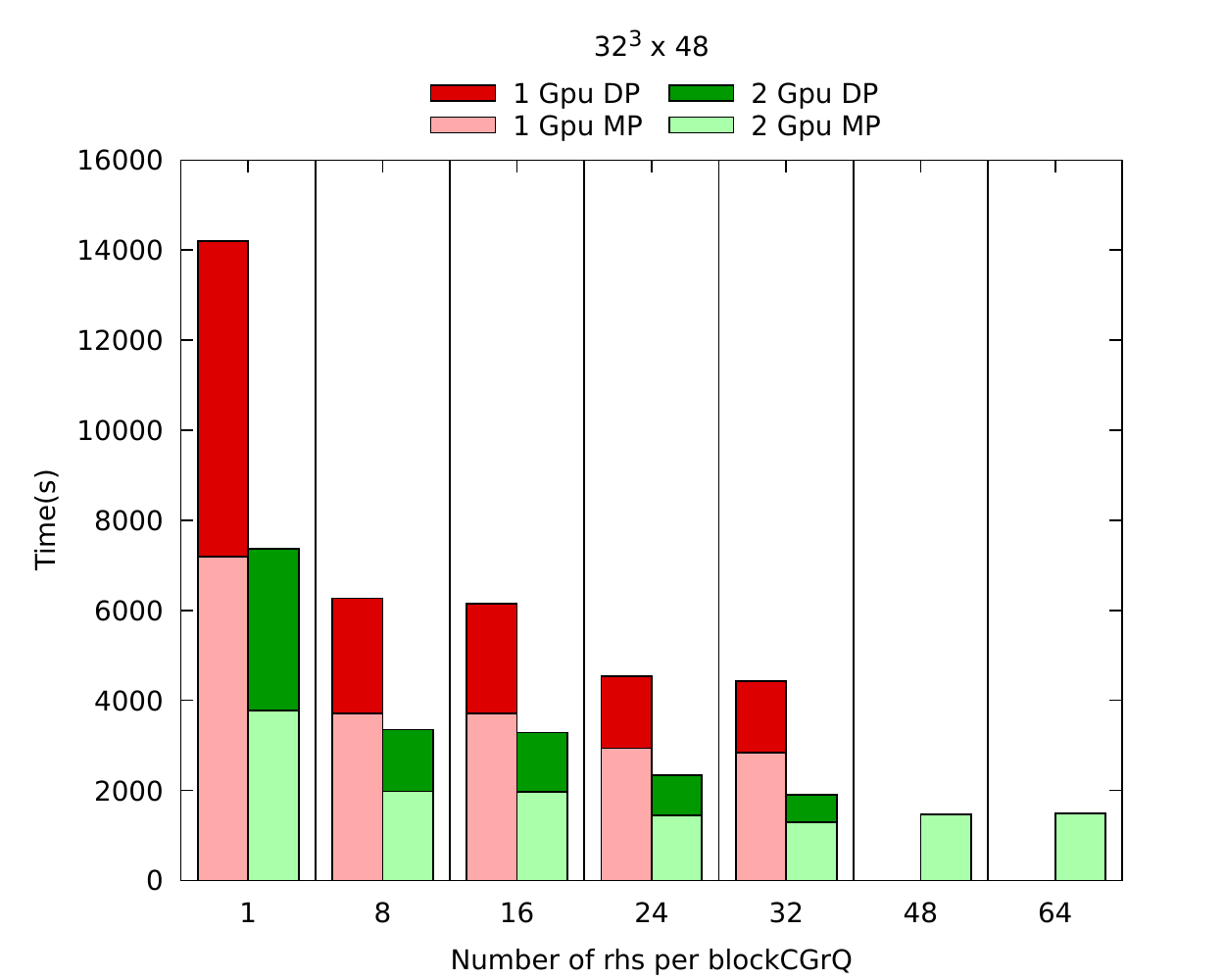}
\caption{\label{fig:tSolSN}Time to solution for the $16^3\times 48$ volume, top, and $32^3 \times 48$ volume, bottom, as a function of the number of RHS. For a fixed number of RHS, the number of GPUs increases from left to right. When the number of RHS is large enough, the system doesn't fit in one GPU and only two GPU results are shown. The two GPU cases for 48 and 64 rhs show a consistent time to solution value between double and mixed-precision.}
\end{figure}

\subsection{Results: Single Node}

In the single-node case, given the optimizations discussed above, our implementation of blockCGrQ should generally display a monotonic decrease in the complete time to solution for all 192 rhs. An appreciable increase in the time to solution would indicate a remnant quadratic scaling in memory streaming, or, in the case of mixed precision, a breakdown of stability. For smaller block sizes, this is in part due to the reduced total iteration counts due to the sharing of the Krylov space inherent to blockCGrQ, but more importantly because we can better saturate memory bandwidth. For larger block sizes, after fully saturating memory bandwidth, the benefit is near solely from the reduced total iteration count.

\begin{figure}[t]
\includegraphics[width=0.95\linewidth,angle=0]{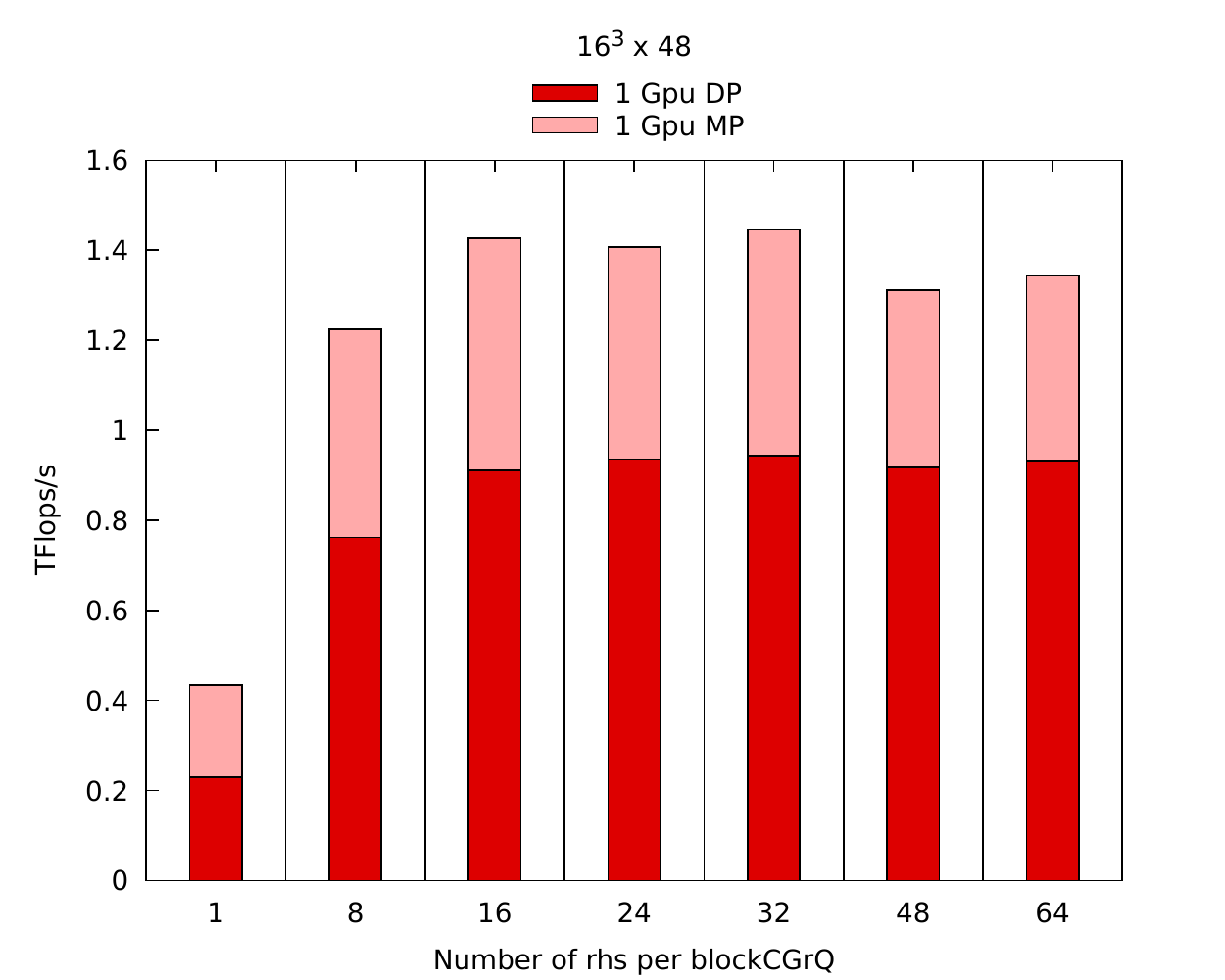}
\\~\\
\includegraphics[width=0.95\linewidth,angle=0]{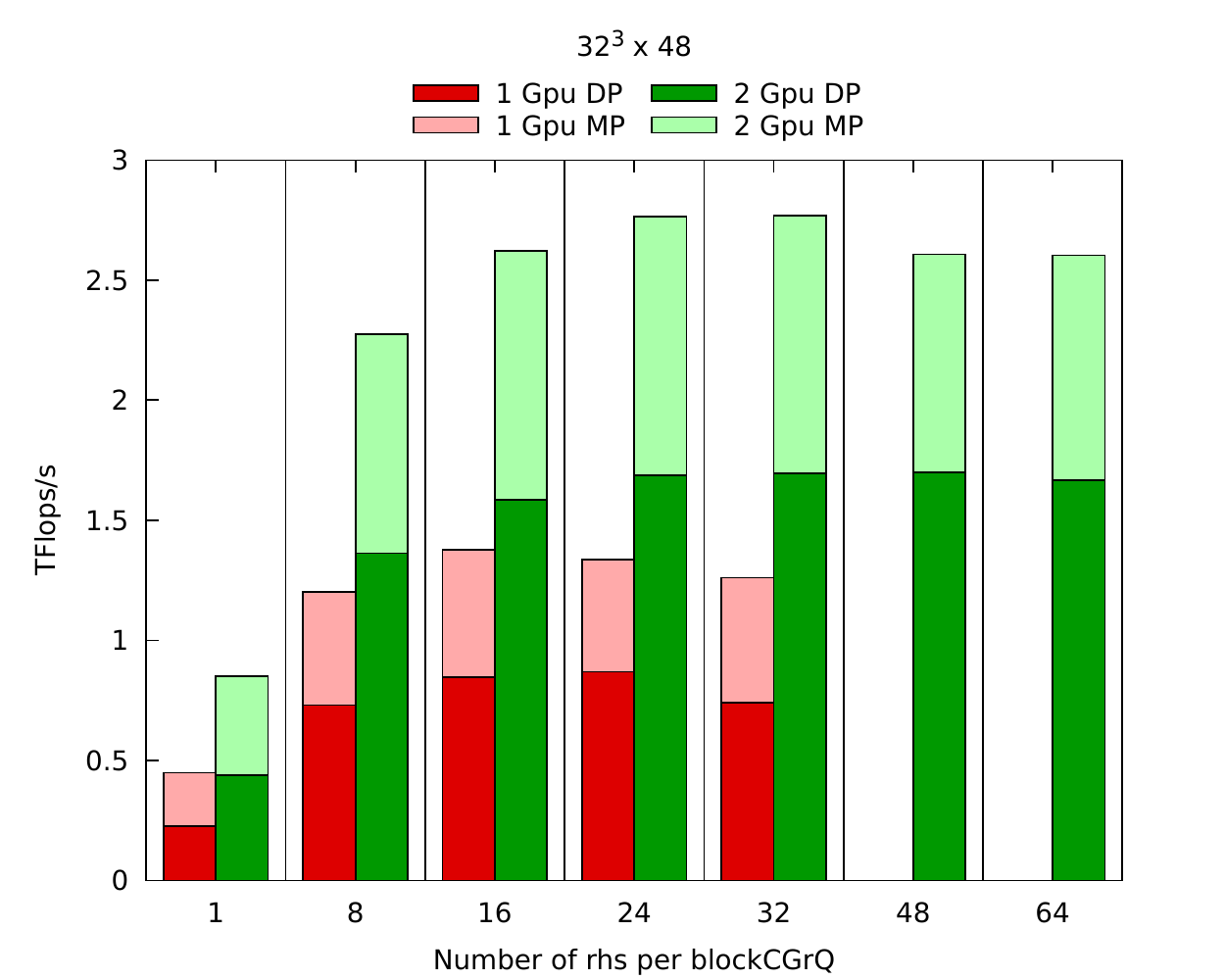}
\caption{\label{fig:TFlopSN}Performance in TFlops/s for the $16^3\times 48$ volume, top, and $32^3 \times 48$ volume, bottom, as a function of the number of rhs.  Notation as in figure~\ref{fig:tSolSN}.}
\end{figure}

This behavior is reflected in figure~\ref{fig:tSolSN}, the total time to solution; figure~\ref{fig:TFlopSN}, the average TFLOPS per second; and the green (lowest) and blue (second highest) curves in \figref{fig:Iters}, the total number of iterations. The {\emph{relative}} reduction in iteration count for a double-precision case is emphasized in \figref{fig:ItersNormed}. We see that the total time to solution does indeed monotonically decrease, reaching a 7$\times$ speedup for double precision and 4$\times$ speedup for mixed-precision, although the benefits become less drastic for a larger block size. The drastic speedup for smaller block sizes is largely due to a better utilization of memory bandwidth, as the iteration count does not greatly decrease. After memory bandwidth has saturated, roughly correlated to a saturation in the TFLOPS per second, the benefit comes from a significant reduction in the number of iterations.

We note that the pure-double-precision solve saturates before the double-single-mixed-precision solve: this reflects both a better utilization of streaming memory bandwidth. For intermediate block sizes, this also reflects a stability of the mixed-precision solve. For larger block sizes, the double-precision solve and the mixed-precision solve take a consistent amount of time (48 and 64). This coincides with a slight breakdown in the stability of the mixed-precision solve, where the reduced time per iteration is offset by the relative increase in iteration count. Regardless, it is still {\emph{slower}} than the mixed-precision solve at a lower block size (32). It is thus important to appropriately tune the block size; bigger is not necessarily better.

\subsection{Results: Strong Scaling}

\begin{figure}[t]
\includegraphics[width=0.95\linewidth,angle=0]{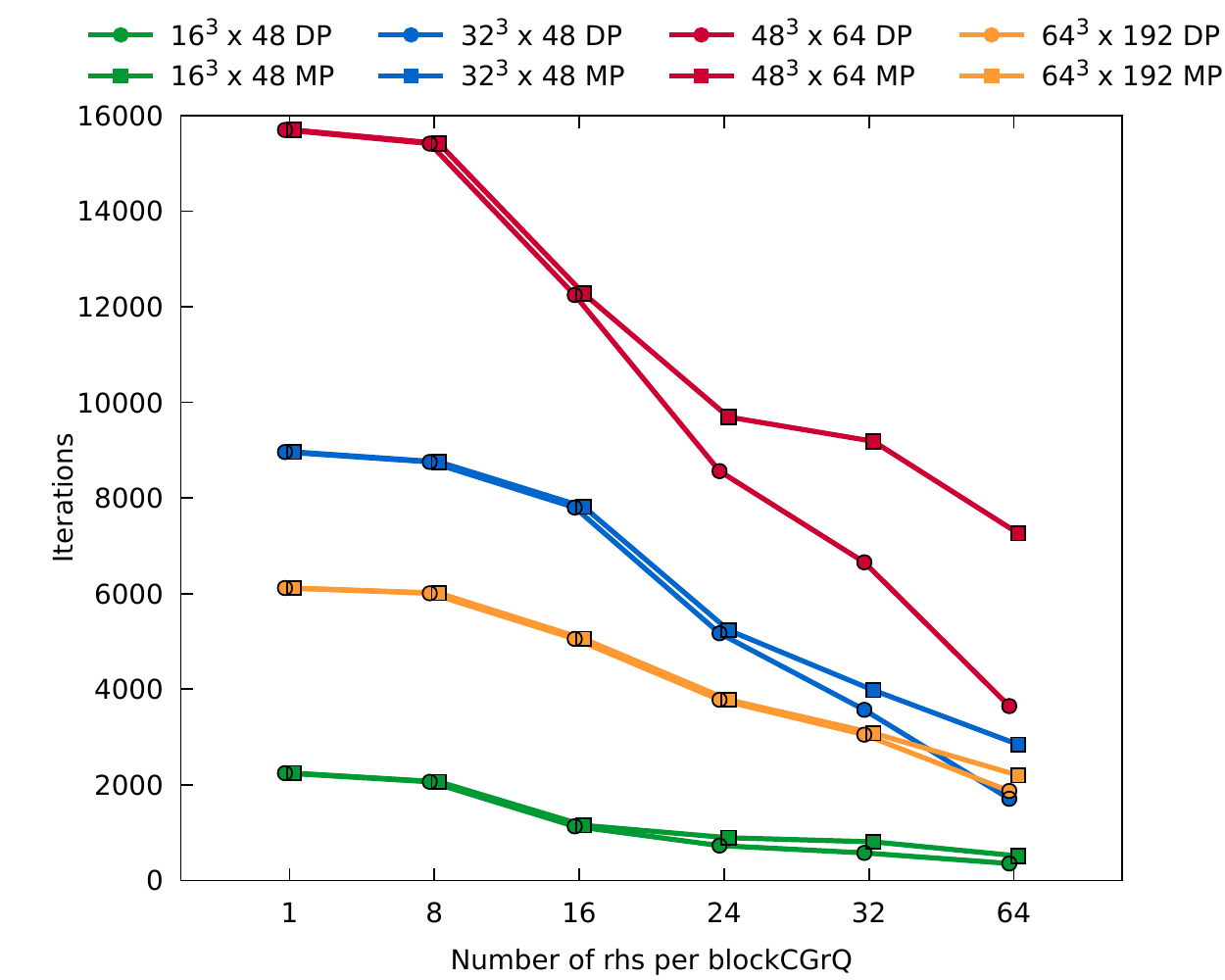}
\caption{\label{fig:Iters}Total iteration count as a function of the number of RHS. The smallest volumes ($16^3\times 48$ and $32^3\times 48$) show the sum of the sloppy + fine iterations,
 whereas the larger volumes only show the fine iterations.}
\end{figure}

\begin{figure}[t!]
\includegraphics[width=0.95\linewidth,angle=0]{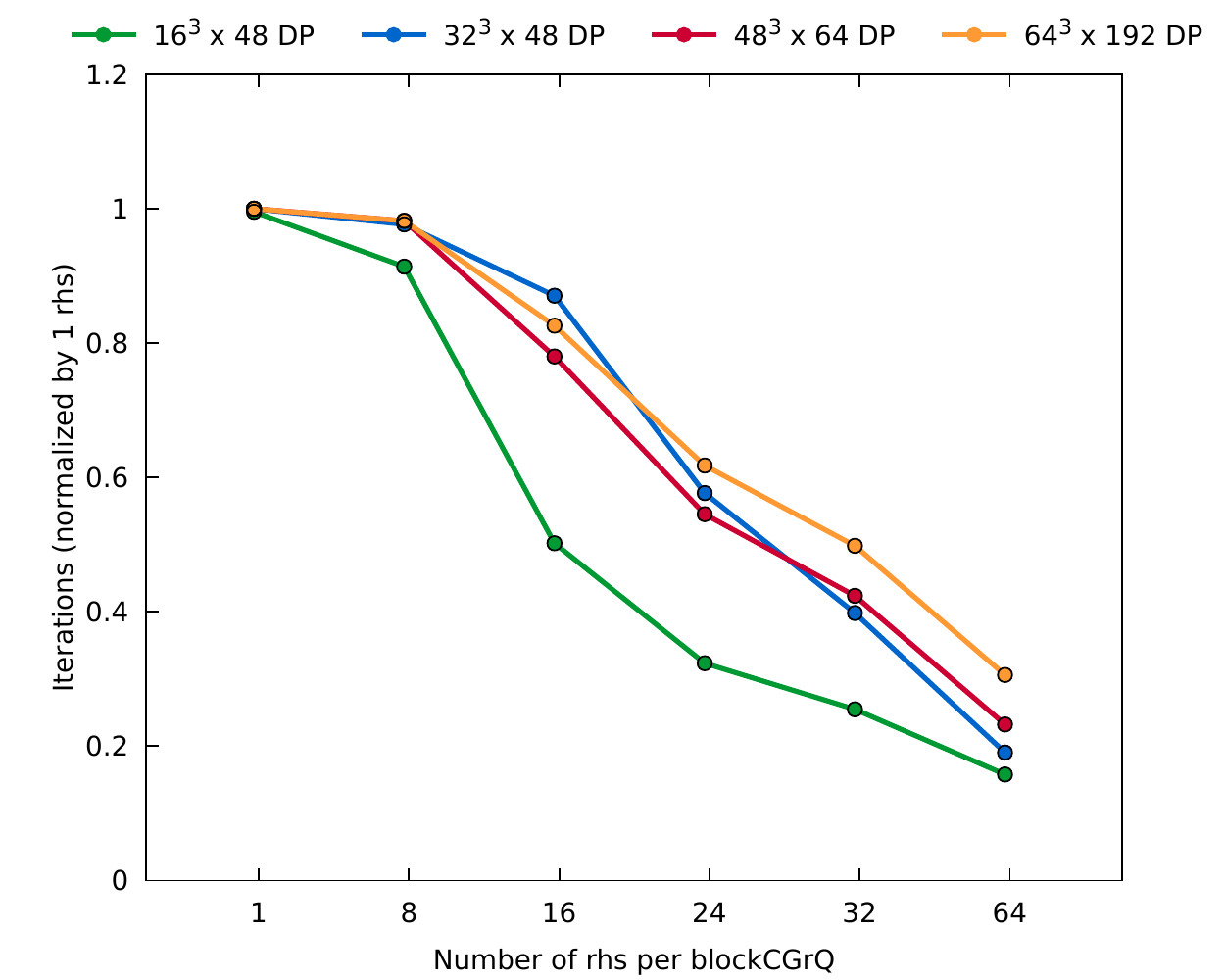}
\caption{\label{fig:ItersNormed}Total iteration count, normalized by the 1 rhs per block case, as a function of the number of RHS. To improve clarity, only the double-precision results
 are shown.}
\end{figure}

\begin{figure}[t!]
\includegraphics[width=0.95\linewidth,angle=0]{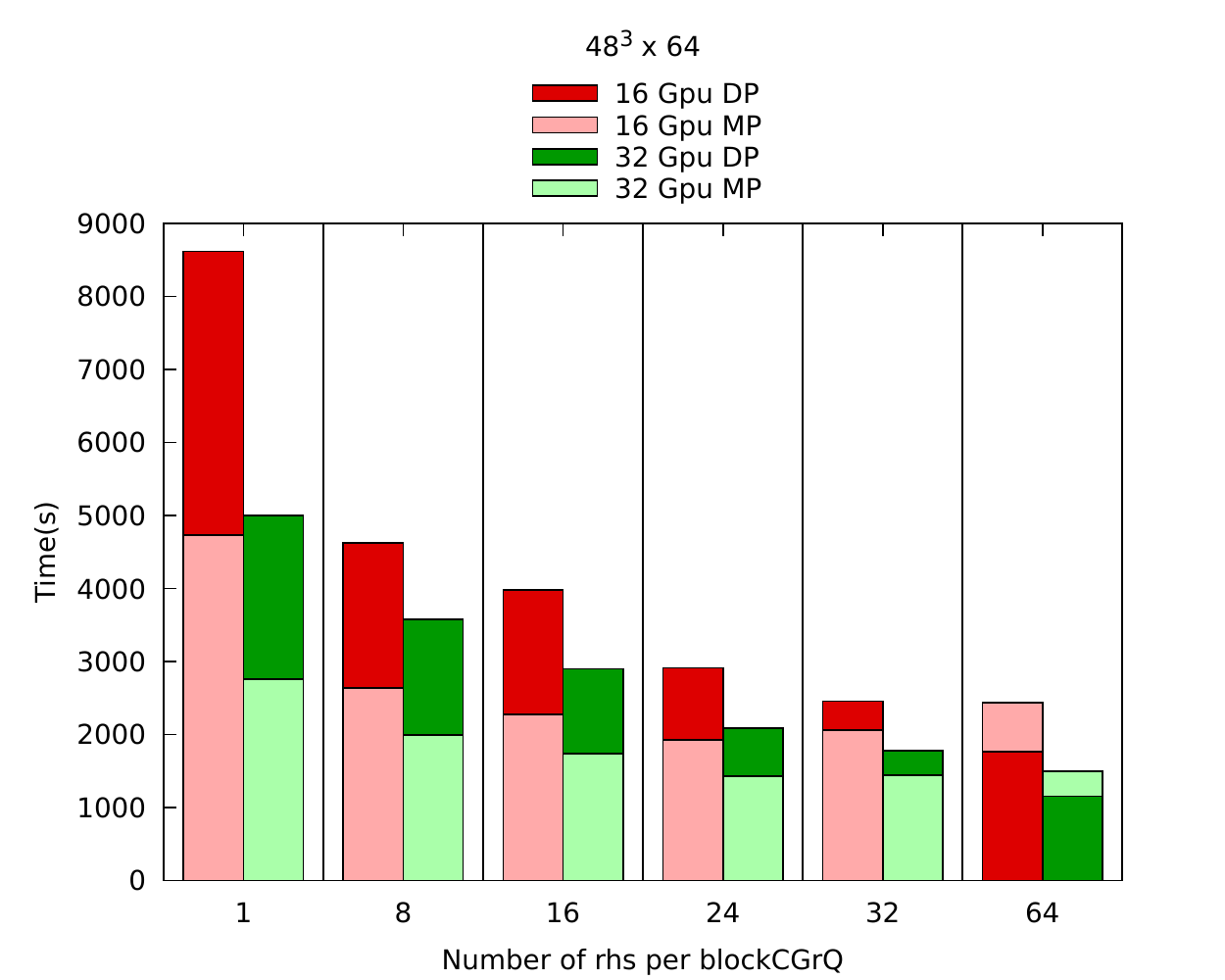}\\~\\\includegraphics[width=0.95\linewidth,angle=0]{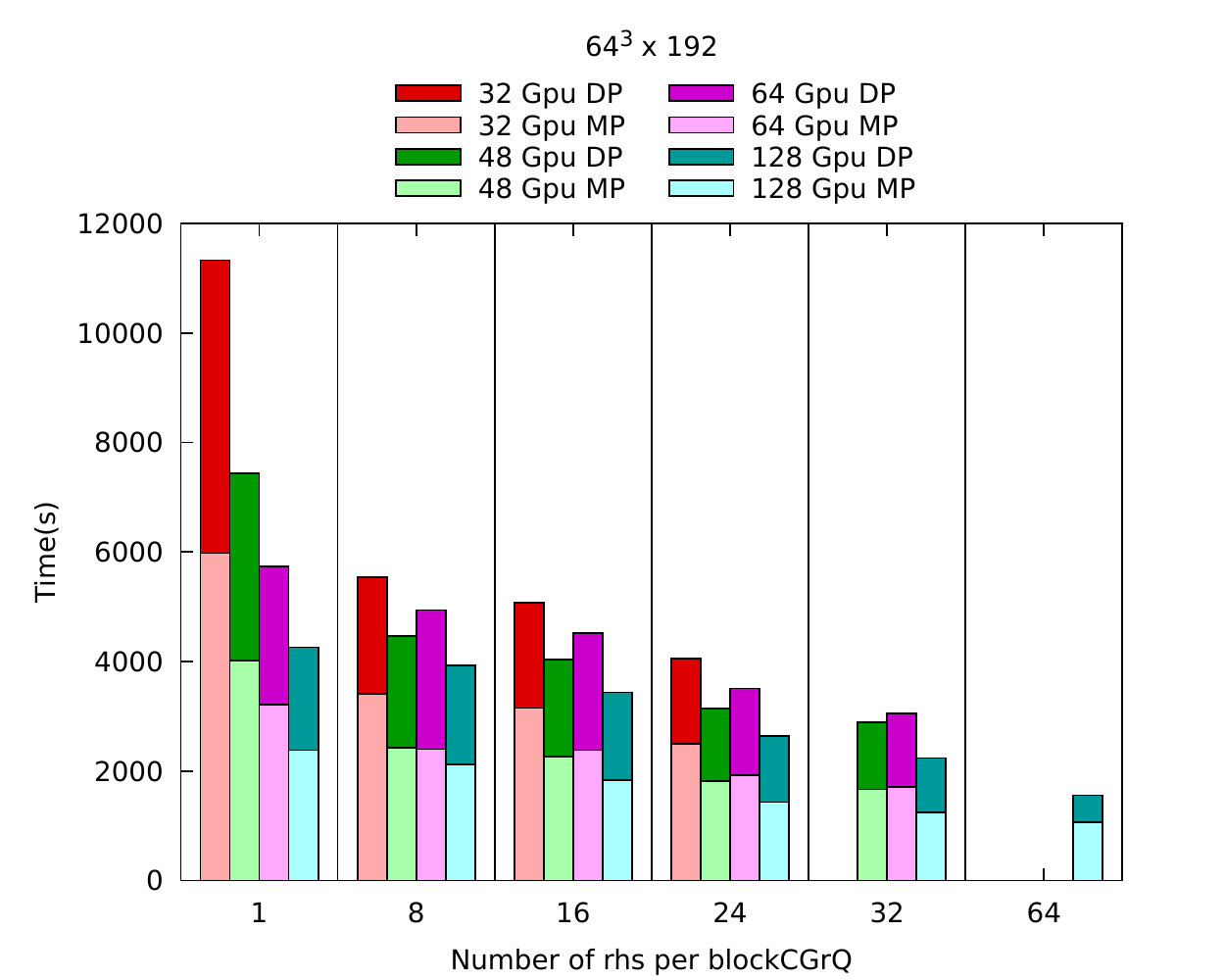}
\caption{\label{fig:tSolMN}Time to solution for the $48^3\times 64$ volume, top, and $64^3 \times 192$ volume, bottom, as a function of the number of RHS. The number of GPUs for a fixed number of RHS grows from left to right.}
\end{figure}

\begin{figure}[t!]
\includegraphics[width=0.95\linewidth,angle=0]{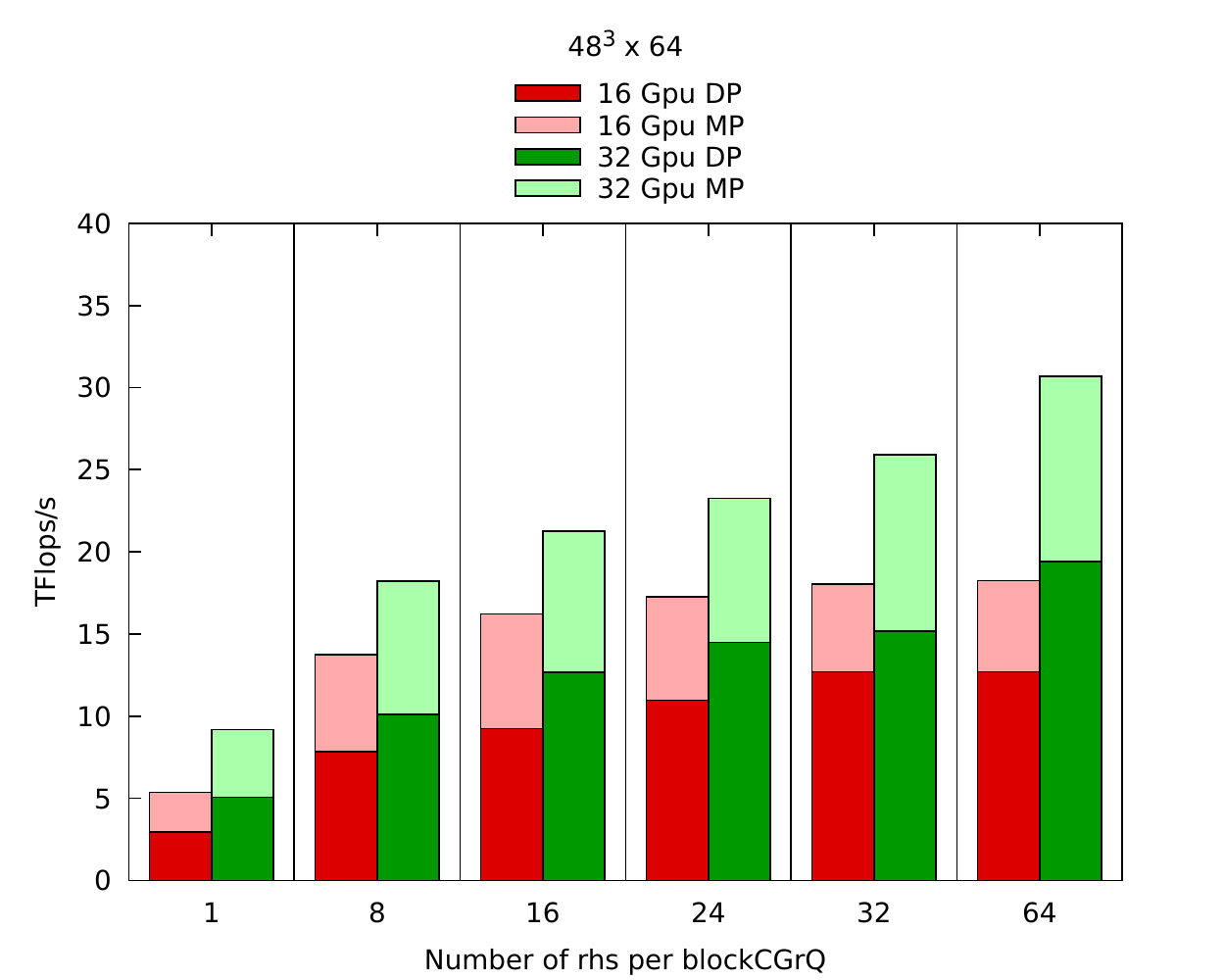}\\~\\\includegraphics[width=0.95\linewidth,angle=0]{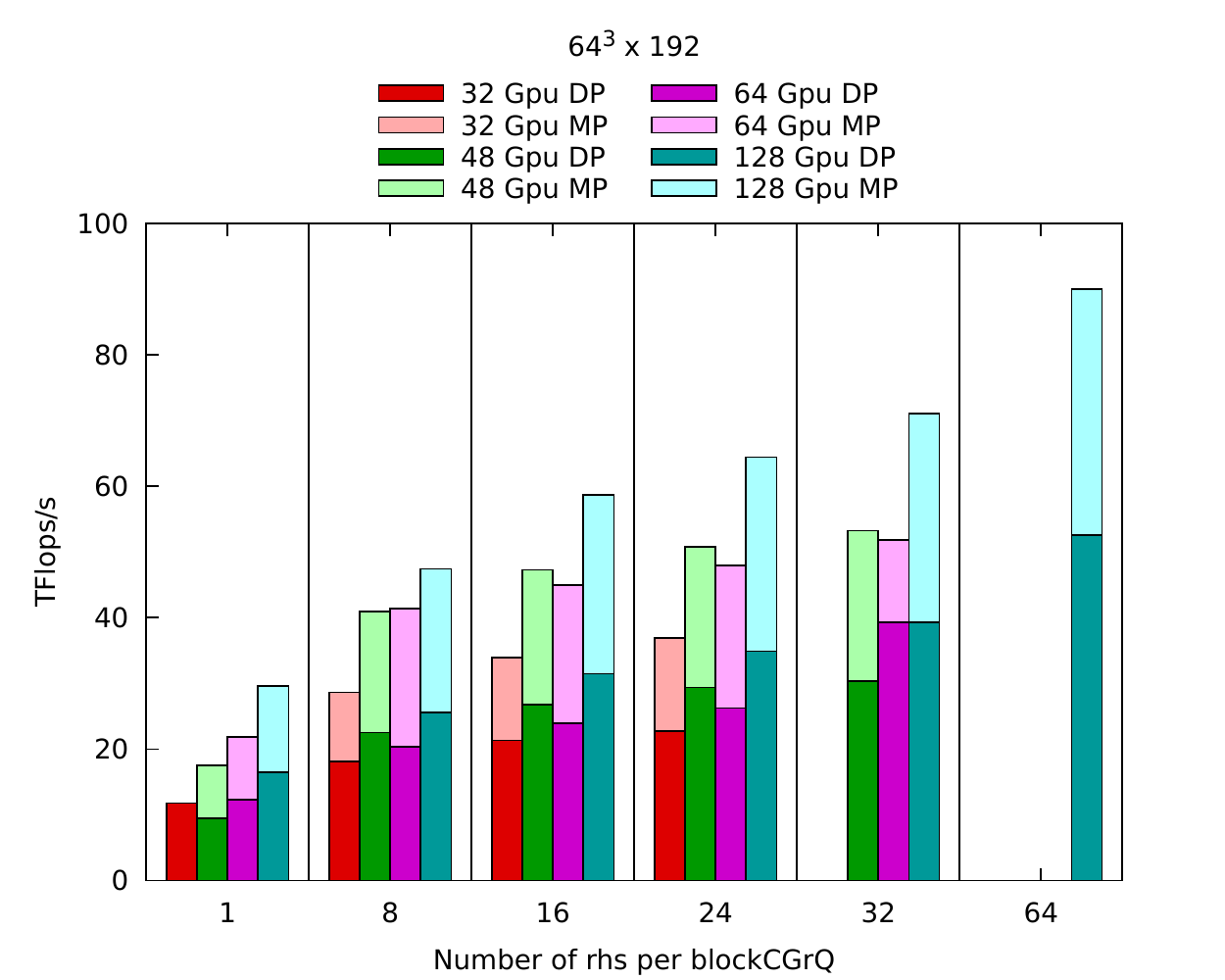}
\caption{\label{fig:TFlopMN}Performance in TFlops/s for the $48^3\times 64$ volume, top, and $64^3 \times 192$ volume, bottom, as a function of the number of RHS. Notation as in figure~\ref{fig:tSolMN}.}
\end{figure}

In considering the strong-scaling regime of our implementation of blockCGrQ, we ran on the NVIDIA SaturnV cluster. The key features to highlight are that we are able to better utilize the GPUs by overlapping computation with network communications during the matrix-vector operation. In addition, we can reduce the overall number of MPI reductions by simultaneously reducing over multiple rhs as per utilizing blockCGrQ. There is also the subtle benefit that, in cases where the local volume per GPU becomes small, we can still saturate streaming memory bandwidth by sufficiently increasing the block size. 

It is difficult to decouple all three of these benefits. However, we can generally comment on the achieved speed-up by again considering figure~\ref{fig:tSolMN}, the total time to solution; figure~\ref{fig:TFlopMN}, the average TFLOPS per second; and the yellow (second lowest) and red (highest) curves in \figref{fig:Iters}, the total number of iterations. The {\emph{relative}} reduction in iteration count for a double-precision case is emphasized in \figref{fig:ItersNormed}.  We emphasize that, despite corresponding to a larger overall volume, it is {\emph{not}} surprising that the X-Large volume (yellow) took far less iterations than the Large volume (red)---the HISQ stencil in the X-Large case is better conditioned than the Large case in one-to-one correspondence with the larger $m_\pi$ in \tabref{tab:mpi}. It is more important to consider the relative behavior {\emph{along}} the curves. 

Our data shows that we achieve an impressive improvement over CG in the strong-scaling regime for a smaller blocksize across both data sets and both precision studies. This is, in some sense, reflective of both the single-node optimizations and the additional strong-scaling benefits. In the case of the X-Large measurement set, the TFLOPS per second monotonically increases with both the number of rhs and the number of GPUs. This may be from a combination of there being increasing work to overlap with communications and additional memory bandwidth to saturate with smaller local volumes (especially in the mixed-precision case). This suggests that, given the resources, the algorithm would continue to strong scale well. The relatively low iterations count corresponds to a relative well-conditioning of the HISQ stencil on the X-Large data set: an increased blocksize always leads to a reduction in the number of iterations.

We see a smaller benefit for the Large measurement set, where in some cases the TFLOPS per second saturates. This could be due to a saturation of memory bandwidth on a local node combined with network latency being hidden behind computation, which in some regards indicates we can strong scale ideally. For very large block sizes there is an increase in the time to solution in the mixed-precision case: this is likely due to a decreased stability in the mixed-precision case. However, in all cases, there is an ideal number of rhs where we can minimize the total time to solution, and for our X-Large data set, that ideal pushes the maximum blocksize and strong-scaling limit we can achieve with our current tests.

In the end the benefits translate into a 2-3$\times$ speedup factor for the mixed double-single precision solver, and a 4-5$\times$ speedup factor for double precision, compared to the baseline.

\section{Future Work}
\label{sec:future}

Our initial implementation of blockCGrQ overcomes many of the
traditional issues which stifled the development and use of block-Krylov
methods. There are several low-hanging fruit that can extend
the application of blockCGrQ in its current form that may be of
interest. For example, deflation methods, or
any other method which improves the initial guess to a linear system,
can be trivially combined with block-Krylov methods.

There are more advanced algorithmic pursuits that can be
considered. As the Lanczos method is closely related to CG, there is
an analogous block-Lanczos method that is related to block CG. At this
time, we do not know if a Lanczos relation can be easily recovered
from the modifications made in blockCGrQ, but if possible it may allow
for the development of a block analogy to the {\emph{EigCG}}
\cite{eigcg} which generates and iteratively improves a
deflation space along the course of a Krylov solve without any
additional stencil applications.

There are further improvements possible within the scope of our
current work. One novel development in this work was the
implementation of reliable updates~\cite{vanderVorst} in a block-Krylov
method. While this was successful in the case of stabilizing a
full double-precision solve as well as a mixed-precision double-single
solve with a modestly sized number of rhs, our implementation of
reliable updates broke down for double-single with a large number of
rhs and completely broke down for a double-half mixed-precision
solve. There are more advanced reliable-update methods in the
literature for non-block-Krylov methods, which could potentially be
generalized for block-Krylov methods. A stable implementation with
half precision would significantly speed up portions of blockCGrQ that
are still memory bandwidth bound. It would also speed up the matrix-vector portions of the algorithm because the HISQ stencil could now be applied in low precision. Last, it would reduce the overall memory footprint which can
be the limiting factor when running at low-node count.

Further improvement to the strong scaling would likely be possible
through utilizing s-step solver methods which reduce the number of
reductions to less than one per solver iteration.  These have been
deployed previously for block solvers to improve parallel
scaling~\cite{doi:10.1002/nla.643}, and such algorithms
could easily be accomodated for in the context of our block-solver
framework.  Given that s-step solvers are typically less numerically
stable, we expect that it will be essential to first develop better
reliable-update methods for blockCGrQ.

As a last note, other non-block-Krylov methods have block-Krylov
extensions. The technology we have developed for an efficient implementation of blockCGrQ, i.e. block-optimized stencil
operations and multi-BLAS operations, are immediately applicable to
other block-Krylov methods.

\section{Conclusions}
\label{sec:conclusion}

In this paper we revisited the blockCGrQ solver proposed
in~\cite{Dubrulle2001}.  We present an implementation that, by
exploiting data locality for reductions and streaming multi-BLAS operation, reduces the na\"ive quadratic scaling of memory traffic to almost linear scaling.
Using a block-optimized matrix-vector operation we
obtain a speedup that multiplicatively combines with the reduced
iteration count from the blockCGrQ algorithm. With its increased
parallelism it is an algorithm very well suited also for future
exascale machines where we expect the gap between peak floating point
operations per seconds and memory bandwidth will continue to increase.
We have also laid the groundwork and demonstrated the algorithm
working using mixed precision, a further important building block for
efficient algorithms. In our real world LQCD use case we
demonstrated a speedup of 5\(\times\), notably not requiring any overhead from
setup and also not requiring large storage for precalculated inputs
such as eigenvectors.  The availability of the building blocks for an efficient implementation
of blockCGrQ can be employed with new algorithms, in order
to improve lattice field theory computational methods to allow new
physics research. Precision results in LQCD are
already competing in accuracy with the best experimental results, but in a lot
of areas computational demands are still the limiting factor.

Although we employed lattice QCD as one use case, we again emphasize that the problem
of efficiently solving a sparse matrix linear system over multiple rhs is a quite general one that
appears in many disciplines. The key features of the implementation that make our BlockCGrQ solver efficient are generic and transfering them to other applications or libraries is expected to be straightforward. 



\section*{Acknowledgement}

The work by A.S. was done for Fermi Research Alliance, LLC under Contract No. DE-AC02-07CH11359 with the U.S. Department of Energy, Office of Science, Office of High Energy Physics.
A.V. was supported by the U.S. National Science Foundation under grant PHY14-14614.
E.W. was supported by the Exascale Computing Project (17-SC-20-SC),
a collaborative effort of the U.S. Department of Energy Office of Science
and the National Nuclear Security Administration.
The authors are grateful for the ORNL-sponsored 2017 GPU Hackathon hosted at the Juelich Supercomputing Center where significant development work was accomplished.

\section*{References}
\bibliographystyle{elsarticle-num}
\bibliography{blocksolver} 

\end{document}